\documentclass[longauth]{aa}  
\usepackage{graphicx}
\usepackage{longtable}
\usepackage{deluxetable}
\usepackage{hyperref}
\usepackage{txfonts}
\begin{document}

   \title{EWOCS-I: The catalog of X-ray sources in Westerlund 1 from the Extended Westerlund 1 and 2 Open Clusters Survey\thanks{Table A.1 is only available in electronic form at the CDS via anonymous ftp to cdsarc.u-strasbg.fr (130.79.128.5) or via http://cdsweb.u-strasbg.fr/cgi-bin/qcat?J/A+A/.}}
%
%   \subtitle{I. Overviewing the $\kappa$-mechanism}

   \author{M. G. Guarcello\inst{1}
          \and
          E. Flaccomio\inst{1}
          \and
          J. F. Albacete-Colombo\inst{2}
          \and
          V. Almendros-Abad\inst{1}
          \and
          K. Anastasopoulou\inst{1}
          \and
          M. Andersen\inst{4}
          \and
          C. Argiroffi\inst{1,5}
          \and
          A. Bayo\inst{4}
          \and
          E. S. Bartlett\inst{4}
          \and
          N. Bastian\inst{6,7,8}
          \and
          M. De Becker\inst{9}
          \and
          W. Best\inst{36}
          \and
          R. Bonito\inst{1}
          \and
          A. Borghese\inst{10,11}
          \and
          D. Calzetti\inst{33}
          \and
          R. Castellanos\inst{12}
          \and
          C. Cecchi-Pestellini\inst{1}
          \and
          J. S. Clark\inst{13}
          \and
          C. J. Clarke\inst{14}
          \and
          F. Coti Zelati\inst{15,16}
          \and
          F. Damiani\inst{1}
          \and
          J. J. Drake\inst{17}
          \and
          M. Gennaro\inst{26,35}
          \and
          A. Ginsburg\inst{18}
          \and
          E. K. Grebel\inst{39}
          \and
          J. L. Hora\inst{17}
          \and
          G. L. Israel\inst{19}
          \and
          G. Lawrence\inst{37,38}
          \and
          D. Locci\inst{1}
          \and
          M. Mapelli\inst{20,21}
          \and
          J. R. Martinez-Galarza\inst{17}
          \and
          G. Micela\inst{1}
          \and
          M. Miceli\inst{1,5}
          \and
          E. Moraux\inst{22}
          \and
          K. Muzic\inst{3}
          \and
          F. Najarro\inst{12}
          \and
          I. Negueruela\inst{23}
          \and
          A. Nota\inst{26}
          \and
          C. Pallanca\inst{24,25}
          \and
          L. Prisinzano\inst{1}
          \and
          B. Ritchie\inst{13}
          \and
          M. Robberto\inst{26,32}
          \and
          T. Rom\inst{22,31}
          \and
          E. Sabbi\inst{26}
          \and
          A. Scholz\inst{27}
          \and
          S. Sciortino\inst{1}
          \and
          C. Trigilio\inst{28}
          \and
          G. Umana\inst{28}
          \and
          A. Winter\inst{29}
          \and
          N. J. Wright\inst{30}
          \and
          P. Zeidler\inst{34}
%\fnmsep\thanks{}
          }

   \institute{Istituto Nazionale di Astrofisica (INAF) – Osservatorio Astronomico di Palermo, Piazza del Parlamento 1, 90134 Palermo, Italy\\
              \email{mario.guarcello@inaf.it}\\
         \and
        Universidad de Rio Negro, Sede Atlántica - CONICET, Viedma CP8500, Río Negro, Argentina \\
        \and
        Instituto de Astrofísica e Ciências do Espaço, Faculdade de Ciências, Universidade de Lisboa, Ed. C8, Campo Grande, 1749-016 Lisbon, Portugal  \\
        \and
        European Southern Observatory, Karl-Schwarzschild-Strasse 2, D-85748 Garching bei M\"{u}nchen, Germany \\
        \and
        Department of Physics and Chemistry, University of Palermo, Palermo, Italy\\
        \and
        Donostia International Physics Center (DIPC), Paseo Manuel de Lardizabal, 4, E-20018 Donostia-San Sebasti\'{a}n, Guipuzkoa, Spain; \\
        \and
        IKERBASQUE, Basque Foundation for Science, E-48013 Bilbao, Spain; \\
        \and
        Astrophysics Research Institute, Liverpool John Moores University, IC2 Liverpool Science Park, 146 Brownlow Hill, Liverpool L3 5RF, UK \\
        \and
        Space Sciences, Technologies and Astrophysics Research (STAR) Institute, University of Liège, Quartier Agora, 19c, Allée du 6 Aôut, B5c, B-4000 Sart Tilman, Belgium \\
        \and
        Instituto de Astrofísica de Canarias, E-38205 La Laguna, Tenerife, Spain \\
        \and
        Departamento de Astrofísica, Universidad de La Laguna, E-38206 La Laguna, Tenerife, Spain \\
        \and
         Departamento de Astrofísica, Centro de Astrobiología, (CSIC-INTA), Ctra. Torrejón a Ajalvir, km 4, Torrejón de Ardoz, E-28850 Madrid, Spain \\
        \and
        School of Physical Sciences, The Open University, Walton Hall, Milton Keynes MK7 6AA, UK \\
        \and
        Institute of Astronomy, University of Cambridge, Madingley Road, Cambridge CB3 0HA, UK \\
        \and
        Institute of Space Sciences (ICE, CSIC), Campus UAB, Carrer de Can Magrans s/n, E-08193, Barcelona, Spain\\ 
        \and
        Institut d'Estudis Espacials de Catalunya (IEEC), Carrer Gran Capit\'{a} 2-4, E-08034 Barcelona, Spain\\
        \and     
        Center for Astrophysics $|$ Harvard \& Smithsonian, 60 Garden Street, Cambridge, MA 02138, USA \\
        \and
        Department of Astronomy, University of Florida, P.O. Box 112055, Gainesville, FL 32611-2055, USA \\
        \and
        Istituto Nazionale di Astrofisica (INAF) - Osservatorio Astronomico di Roma, Via Frascati 33, I-00078 Monte Porzio Catone, Italy\\
       \and
       Universit\"at Heidelberg, Zentrum f\"ur Astronomie, Institut f\"ur Theoretische Astrophysik, Albert-Ueberle-Str. 2, D--69120 Heidelberg, Germany\\
       \and
        Physics and Astronomy Department Galileo Galilei, University of Padova, Vicolo dell'Osservatorio 3, I-35122 Padova, Italy \\
        %\and
        %Istituto Nazionale di Astrofisica (INAF) – Osservatorio Astronomico di Padova, Vicolo dell'Osservatorio 5, I-35122 Padova, Italy \\
        \and
        Univ. Grenoble Alpes, CNRS, IPAG, 38000 Grenoble, France \\
        \and
        Departamento de Física Aplicada, Facultad de Ciencias, Universidad de Alicante, Carretera de San Vicente s/n, E-03690, San Vicente del Raspeig, Spain \\
        \and
        Dipartimento di Fisica e Astronomia, Università di Bologna, Via Gobetti 93/2, Bologna I-40129, Italy; \\
        \and
        Istituto Nazionale di Astrofisica (INAF) - Osservatorio di Astrofisica e Scienza dello Spazio di Bologna, Via Gobetti 93/3, Bologna I-40129, Italy \\
        \and
        Space Telescope Science Institute, 3700 San Martin Dr, Baltimore, MD, 21218, USA \\
        \and
        School of Physics \& Astronomy, University of St Andrews, North Haugh, St Andrews KY16 9SS, UK \\
        \and
        Istituto Nazionale di Astrofisica (INAF) – Osservatorio Astrofisico di Catania, Via Santa Sofia 78, I-95123 Catania, Italy \\
        \and
         Universit\'{e} C\^{o}te d'Azur, Observatoire de la C\^{o}te d'Azur, CNRS, Laboratoire Lagrange, F-06300 Nice, France; Universit\'{e} Grenoble Alpes, CNRS, IPAG, F-38000 Grenoble, France\\
        \and
        Astrophysics Group, Keele University, Keele, Staffordshire ST5 5BG, United Kingdom\\
        \and
        University of Split, Faculty of Science, Department of Physics, Rudera Boškovića 33, 21000, Split, Croatia \\
        \and
        Johns Hopkins University, 3400 N. Charles Street, Baltimore, MD 21218, USA \\
        \and
        Department of Astronomy, University of Massachusetts, 710 North Pleasant Street, Amherst, MA 01003, USA \\
        \and
        AURA for the European Space Agency (ESA), ESA Office, Space Telescope Science Institute, 3700 San Martin Drive, Baltimore, MD 21218, USA \\
        \and
        The William H. Miller III Department of Physics \& Astronomy, Bloomberg Center for Physics and Astronomy, Johns Hopkins University, 3400 N. Charles Street, Baltimore, MD 21218, USA \\
        \and
        Department of Astronomy, University of Texas at Austin, Austin, TX 78712, USA \\
        \and
        Centre for Astrophysics and Supercomputing, Swinburne University of Technology, PO Box 218, Hawthorn VIC 3122, Australia \\
        \and
        Department of Physics \& Astronomy, University College London, Gower Street, London WC1E 6BT, UK \\
        \and
        Astronomisches Rechen-Institut, Zentrum f\"ur Astronomie der Universit\"at Heidelberg, M\"onchhofstr.\ 12--14, 69120 Heidelberg, Germany
        }
   \date{}
 
  \abstract
   {With a mass exceeding several $10^4$ M$_\odot$ and a rich and dense population of massive stars, supermassive young star clusters represent the most massive star-forming environment that is dominated by the feedback from massive stars and gravitational interactions among stars.}
   {In this paper we present the "Extended Westerlund 1 and 2 Open Clusters Survey" (EWOCS) project, which aims to investigate the influence of the starburst environment on the formation of stars and planets, and on the evolution of both low and high mass stars. The primary targets of this project are Westerlund~1 and 2, the closest supermassive star clusters to the Sun.}
   {The project is based primarily on recent observations conducted with the \emph{Chandra} and JWST observatories. Specifically, the \emph{Chandra} survey of Westerlund~1 consists of 36 new ACIS-I observations, nearly co-pointed, for a total exposure time of 1 Msec. Additionally, we included 8~archival \emph{Chandra}/ACIS-S observations. This paper presents the resulting catalog of X-ray sources within and around Westerlund~1. Sources were detected by combining various existing methods, and photon extraction and source validation were carried out using the \textit{ACIS}-Extract software.}
   {The EWOCS X-ray catalog comprises 5963 validated sources out of the 9420 initially provided to \textit{ACIS}-Extract, reaching a photon flux threshold of approximately $2\times10^{-8}$ photons$\,$cm$^{-2}\,$s$^{-1}$. The X-ray sources exhibit a highly concentrated spatial distribution, with 1075 sources located within the central 1 arcminute. We have successfully detected X-ray emissions from 126 out of the 166 known massive stars of the cluster, and we have collected over 71000 photons from the magnetar CXO J164710.20-455217.}
   {}


   \maketitle
%
%-------------------------------------------------------------------
\par$ $
\section{Introduction}

The star formation rate and the properties of the most common star-forming environments in galaxies vary over time. When considering cosmological timescales, the star formation rate is known to reach its peak at approximately z$\sim$2--3 and then gradually decline \citep[e.g.,][]{HopkinsBeacom2006ApJ...651..142H,Dunlop2011Sci...333..178D,MadauDickinson2014ARAA..52..415M}. In the local Universe, mergers play a dominant role in shaping the star formation process in galaxies \citep{RiekeRujopakarn2011ASPC..446....3R} as they influence the overall properties of the interstellar medium. Such interactions happen frequently, and various studies have demonstrated that interacting galaxies undergo periods of intense star formation \citep[e.g.,][]{LarsonTinsley1978ApJ...219...46L,Smith2010AJ....140.1975S} because of the considerable impact interactions have on the stars formation process, for instance from the enhancement of the star cluster formation rate due to close encounters (e.g. in the Magellanic Clouds) and the ram pressure stripping enhancing star formation (e.g., jellyfish galaxies). Noticeable examples are the nearby interacting galaxies M51 and M82, where we observe extreme star formation taking place in very massive young clusters with masses reaching several times $10^5\,$M$_{\odot}$ \citep[known as super star clusters;][]{deGrijs2001AJ....121..768D,deGrijs2003MNRAS.342..259D,deGrijs2003MNRAS.343.1285D}. Generally, these highly massive star clusters constitute the dominant star-forming environments in starburst galaxies and are likely prevalent during the peak era of cosmic star formation \citep[e.g.,][]{Figer2008IAUS..250..247F,Adamo2020SSRv..216...69A}.\par

In the Milky Way, current estimates of the star forming heavily rely on the methods employed. For example, \citet{RobitailleWhitney2010ApJ...710L..11R} derived a range of 0.68--1.45 M${_\odot}\,$yr$^{-1}$ based on the population of young stellar objects identified in the \emph{Spitzer}/IRAC survey of the Galactic plane GLIMPSE \citep{Benjamin2003PASP..115..953B}. On the other hand, \citet{Licquia2015ApJ...806...96L} applied a hierarchical Bayesian statistical method to previous analyses and determined a star formation rate of about 1.6$\,$M${_\odot}$/yr. For comparison, recent estimates of the star formation rate in M51 range from 4.8$\,$M${_\odot}\,$yr$^{-1}$ \citep[from a 158$\mu m$ map of the galaxy][]{Pineda2018ApJ...869L..30P} to 2.7$\,$M${_\odot}\,$yr$^{-1}$ \citep[from combined UV+optical spectral energy distribution fitting;][]{Eufrasio2017ApJ...851...10E}, while in M82 star formation rates of 2-4$\,$M${_\odot}\,$yr$^{-1}$ were observed \citep{deGrijs2001AJ....121..768D}. Nevertheless, all these studies indicate that our Galaxy does not currently have a high star formation rate. Consequently, it is not surprising that the Milky Way lacks a prominent population of super star clusters with masses exceeding 10$^4$ M$_{\odot}$. In order of distance from the Sun, the most massive clusters known are Westerlund~1 (2.6-5 kpc; \citealp{Aghakhanloo2020MNRAS.492.2497A}, \citealp{Clark2005A&A...434..949C}), Westerlund 2 ($\sim$4.2 kpc; \citealp{VargasAlvarez2013AJ.145.125V}), NGC 3603 (7.6 kpc; \citealp{Melena2008AJ....135..878M}), the Arches and Quintuplet clusters (both at $\sim$8.5 kpc; \citealp{Figer2002ApJ...581..258F}, \citealp{Figer1999ApJ...514..202F}), Mercer~81 (11 kpc; \citealp{Davies2012MNRAS.419.1860D}), and Mercer~30 (12 kpc; \citealp{DelaFuente2016AA...589A..69D}). Similar regions in terms of mass, but with with a low stellar density, are the Cygnus~OB2 association (1.4 kpc; \citealp{RyglBSM2012}) and the W3 complex (about 2 kpc; \citealp{Hachisuka2006ApJ...645..337H}). Slightly older supermassive star clusters (10-20 Myrs) are found in the Scutum-Crux arm (about 6 kpc from the Sun; \citealp{Figer2006ApJ...643.1166F}; \citealp{Davies2007ApJ...671..781D}; \citealp{Clark2009AA...498..109C}). Despite their limited number, these super star clusters hold significant importance as they enable the study of star and planet formation, as well as early stellar evolution, in a star-forming environment that was characteristic of epochs when the Milky Way had higher rates of star formation than today and most of the field stars in our Galaxy formed. \par

In this paper we present the Extended Westerlund 1 and 2 Open Clusters Survey (EWOCS) project, which is focused on studying star and planet formation and early stellar evolution in compact starbursts, using Westerlund~1 and 2 as first science cases. In particular, this paper focuses on the catalog of X-ray sources detected in the deep \emph{Chandra} observations of Westerlund~1. The paper is organized as follows: We present Westerlund~1 and the EWOCS project in Sect. \ref{sect_Wd1}. The EWOCS observations are described in Sect. \ref{sec_data}, the procedure for source detection is described in Sect. \ref{sec_detection} and that of source validation and extraction in Sect. \ref{sec_extraction}. The final catalog of the X-ray sources in Westerlund~1 is described in Sect. \ref{sec_finalcatalog}. \par

%--------------------------------------------------------------------
\section{Westerlund~1 and the EWOCS project}
\label{sect_Wd1}

Westerlund~1 is located at RA$\rm_{J2000}$=16h47m04s and dec.$\rm_{J2000}$=-45$^\circ$51$^\prime$05$^{\prime\prime}$, corresponding to Galactic coordinates l=339.55$^\circ$ and b=-00.40$^\circ$. The cluster was discovered by \citet{Westerlund1961PASP.73.51W} through observations made with the 26-inch Uppsala-Schmidt telescope at Mt. Stromlo Observatory in Australia. From these initial observations, it became evident that Westerlund~1 is a very massive cluster. Today, it is considered to be the most massive young cluster known within the Milky Way, with mass estimates ranging from approximately 5$\times$10$^4$ M${_\odot}$ to over 10$^5$ M${_\odot}$ \citep{Clark2005A&A...434..949C,Brandner2008AA.478.137B,Gennaro2011,Lim2013AJ....145...46L,Andersen2017AA.602A.22A}.

Despite over 60 years of studies and observations of Westerlund~1, the tension regarding the parameters of this distinctive cluster remains unresolved. This is primarily due to its compact nature and the significant extinction that has long hindered the ability to resolve its low-mass stars. \par

The distance to the cluster has been a subject of long-standing debate. The initial estimate by \citet{Westerlund1961PASP.73.51W} was of 1.4$\,$kpc. However, the same authors later presented a more distant estimate of 5$\,$kpc based on photographic observations in the $VRI$ bands \citep{Westerlund1968ApJ...154L..67W}. The first study utilizing CCD imaging of the cluster \citep{Piatti1998AAS..127..423P} reported a distance estimate of 1.0$\pm$0.4$\,$kpc. However, this estimate was based on the incorrect assumption that all cluster members were on the main sequence.\par

Several authors have made distance estimates for Westerlund~1 based on the analysis of its rich population of massive stars. For example, \citet{Clark2005A&A...434..949C} based their estimate on six yellow hypergiants (YHGs), assuming that these stars have the standard luminosity for this class of objects \citep[log(L/L$_\odot$)$\sim$5.7;][]{Smith2004ApJ...615..475S}, and adopting an extinction of A$\rm_V$=11$^m$, resulting in a distance range between 2 kpc and 5.5 kpc. A similar value was found by \citet{Crowther2006MNRAS.372.1407C} through infrared analysis of WN and WC stars. \citet{KoumpiaBonanos2012AA...547A..30K} derived a distance of 3.7$\pm$0.6 kpc from the analysis of the dynamics and geometry of the eclipsing binary W13. By comparing the cluster locus in color-magnitude diagrams with suitable isochrones, \citet{Brandner2008AA.478.137B} determined a distance of 3.55$\pm$0.17 kpc, while \citet{Gennaro2011} found a distance of 4.0$\pm$0.2 kpc, and \citet{Lim2013AJ....145...46L} reported a distance of about 3.8 kpc. An independent estimate (3.9$\pm$0.7 kpc) was provided by \citet{Kothes2007AA...468..993K} using the radial velocity of HI clouds in the direction of the cluster, assuming they were physically connected to Westerlund~1.\par

More recently, \emph{Gaia} data have been extensively utilized to measure the distances of star clusters, providing precise values up to distances of about 1 kpc \citep{Gaia2016AA595A2G}. However, for more distant clusters, careful analysis and assumptions are required to obtain reliable distance measurements. Consequently, it is not surprising that different estimates of the distance to Westerlund~1 have emerged from authors who have analyzed \emph{Gaia} data. \citet{Aghakhanloo2020MNRAS.492.2497A} conducted a Bayesian analysis of \emph{Gaia} data along the line of sight to Westerlund~1 and obtained a mean cluster parallax of 0.35$^{+0.07}_{-0.06}$ mas, which corresponds to a distance of 2.6$^{+0.6}_{-0.4}$ kpc and is in tension with the previous estimate of approximately 0.19 mas provided by \citet{Clark2020AA...635A.187C}. Focusing on known members of Westerlund~1, \citet{DaviesBeasor2019MNRAS.486L..10D} found a distance of 3.9$^{+1.0}_{-0.64}$ kpc. More recently, \citet{Negueruela2022AA...664A.146N} carried out a detailed determination of candidate members in Westerlund~1 using \emph{Gaia} Early Third Data Release \citep[EDR3;][]{vanLeeuwen2021gdr3.reptE....V} data and obtained a distance of 4.23$^{+0.23}_{-0.21}$ kpc, suggesting that the cluster is located in the Norma arm. A similar estimate from the \emph{Gaia}/EDR3 was obtained by \citet{Navarete2022MNRAS.516.1289N}.\par

Given the uncertainty surrounding the distance to Westerlund~1, it is not surprising that estimates of the cluster's age provided by different authors also vary significantly. Age estimates in the range of 3.2 to 5 million years have been derived using isochrone fitting on the high-mass sequence and arguments based on the diverse population of massive stars, including Wolf-Rayet (WR) stars, YHGs, and red supergiants \citep[RSGs;][]{Clark2005A&A...434..949C,Crowther2006MNRAS.372.1407C,Brandner2008AA.478.137B,Ritchie2010AA...520A..48R,Gennaro2011,KoumpiaBonanos2012AA...547A..30K,Kudryavtseva2012ApJ.750L.44K,Mackey2015AA...582A..24M}. These authors found a relatively narrow age spread, with an upper limit of 0.4 million years indicating that Westerlund~1 likely formed in a single burst of star formation \citep{Kudryavtseva2012ApJ.750L.44K}. However, some of these estimates are based on arguments that strictly apply to single stars, whereas it is known that the binary fraction among the massive members of Westerlund~1 is very high \citep{Crowther2006MNRAS.372.1407C}. More recent studies suggest a more complex star formation history and a slightly older age \citep{Aghakhanloo2020MNRAS.492.2497A, Beasor2021ApJ...912...16B,Navarete2022MNRAS.516.1289N,Negueruela2022AA...664A.146N}. In particular, arguments based on spectral energy distribution fitting and the luminosity of individual RSGs support an age estimate exceeding 10 Myrs \citep{Beasor2021ApJ...912...16B,Navarete2022MNRAS.516.1289N}, although this estimate is in tension with other properties of the cluster.\par

\begin{figure*}[!h]
\centering
\includegraphics[width=0.45\textwidth]{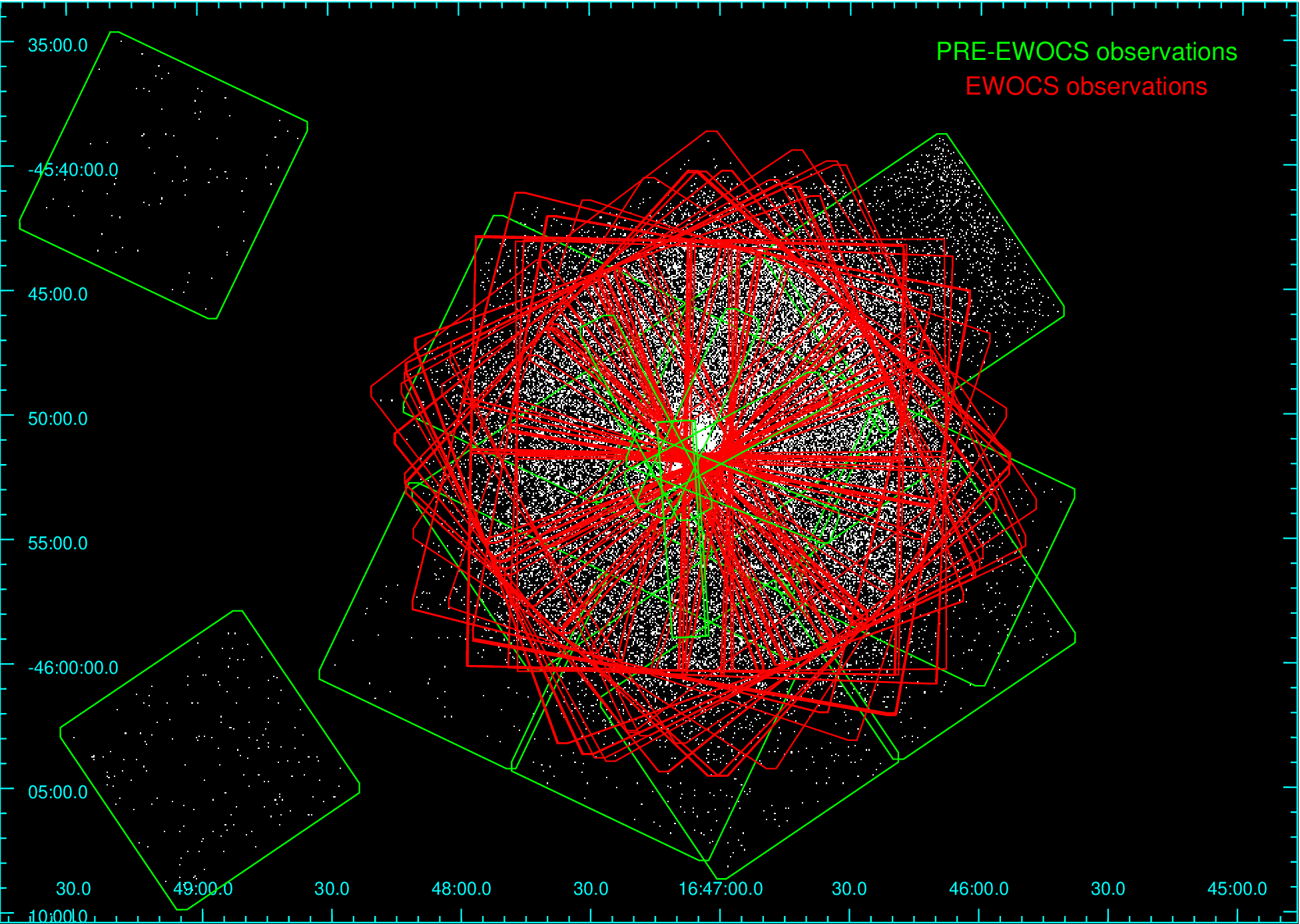}
\includegraphics[width=0.45\textwidth]{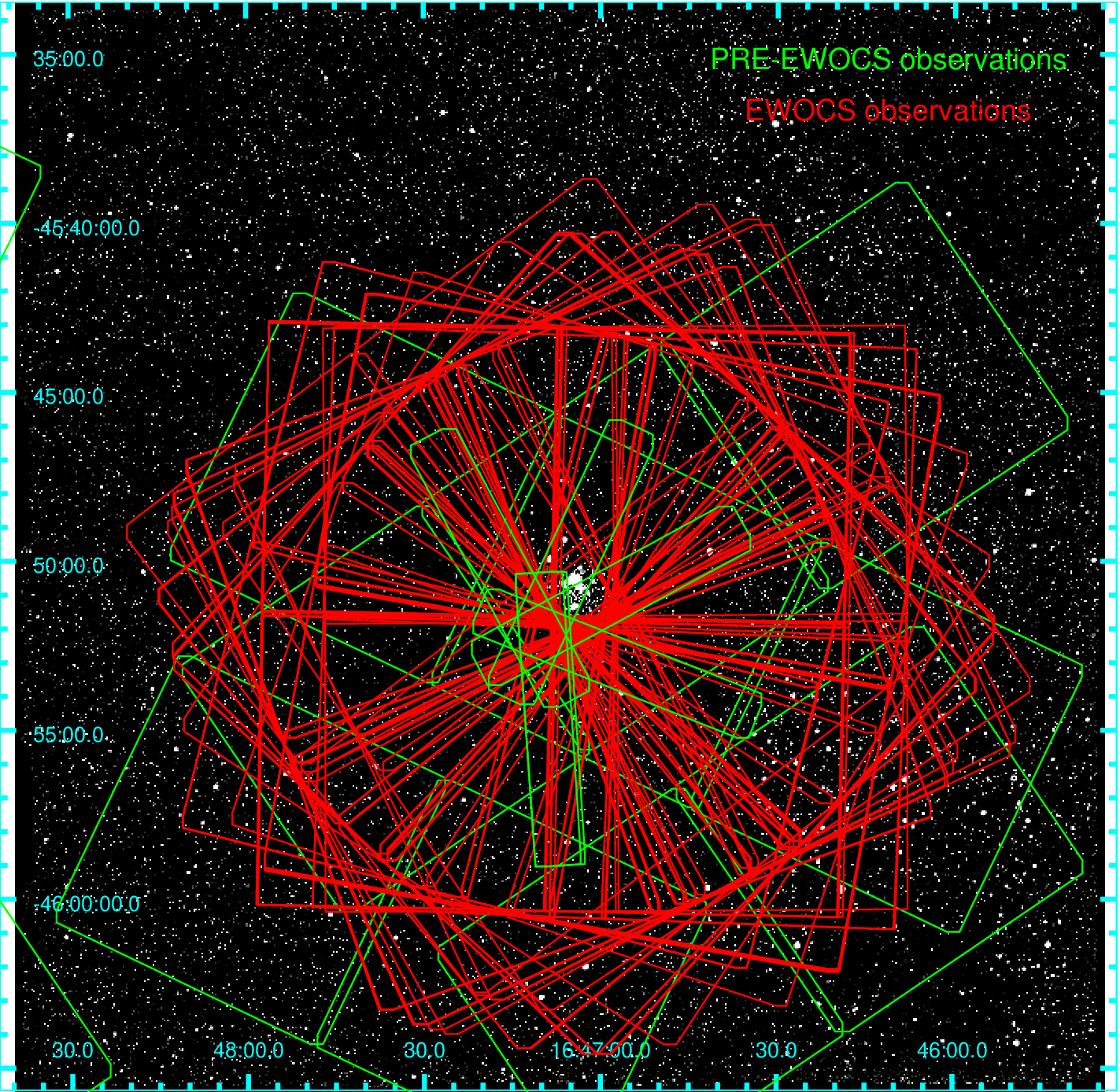}
\caption{Contours of the pre-EWOCS and EWOCS \emph{Chandra} observations of Westerlund~1 overlaid on the combined ACIS event file (left panel) and on an image in the $Ks$ band obtained with the FourStar infrared camera mounted on the Magellan 6.5$\,$m telescopes (right panel).}
\label{fig_evtcontours}%
\end{figure*}

There is a general consensus in the literature regarding other important properties of Westerlund~1, including its high extinction, large mass, and notably, its impressive population of massive stars. The significant extinction toward Westerlund~1 has been acknowledged since the initial publication on this cluster, where an approximate visual extinction of A$_V$$\sim$12$^m$ has been found \citep{Westerlund1961PASP.73.51W}. Subsequent estimates range from 10$^m$ to 13$^m$ of visual extinction \citep{Negueruela2010,Lim2013AJ....145...46L,Damineli2016MNRAS.463.2653D}. There is some disagreement regarding the extinction law in the direction of Westerlund~1: While according to \citet{Negueruela2010} it follows the standard law in the $VRI$ bands, \citet{Lim2013AJ....145...46L} and \citet{Damineli2016MNRAS.463.2653D} suggested a steeper extinction law in the near-IR (R$\rm_V$=2.50$\pm$0.04). \par

The most remarkable characteristic of Westerlund~1 is its large population of massive stars \citep{Clark2005A&A...434..949C,Ritchie2009AA...507.1585R,Clark2020AA...635A.187C}, which includes 24 WR stars \citep{ClarkNegueruela2002AA...396L..25C,NegueruelaClark2005AA...436..541N,Skinner2006ApJ...639L..35S,Groh2006AA...457..591G,Crowther2006MNRAS.372.1407C}, the luminous blue variable (LBV) Wd1-243 \citep{ClarkNegueruela2004AA...413L..15C}, ten YHGs with spectral classes ranging from A5Ia$^+$ to F8Ia$^+$ \citep{Clark2005A&A...434..949C}\footnote{A different classification for six YHGs has recently been presented by \citet{Beasor2023arXiv230316937B}}, two blue stragglers \citep{Clark2019AA...623A..83C}, four RSGs \citep{Wright2014MNRAS.437L...1W}, seven blue hypergiants (BHGs), and over 100 bright OB supergiants dominated by spectral classes O9-B1 \citep{Negueruela2010}. Most of these sources are concentrated in the inner region of the cluster, spanning approximately 1 arcminute, with only a few more isolated massive stars, such as WR77.\par

In particular, Westerlund~1 hosts examples of every known transitional evolutionary phase between H-rich OB supergiants and H-depleted WR stars. This makes the cluster a unique target for studying massive stars and, specifically, for understanding how binarity and mass loss impact the evolutionary paths of these stars and how the initial stellar masses are linked to the types of compact objects that form at the end of their evolution. Winds and mass loss in these stars have been extensively studied with radio and millimeter-continuum observations, which have detected individual bright sources, such as W9, surrounded by extended nebulae, providing evidence of intense mass loss in the past \citep[up to several 10$^{-4}\,$M$_\odot$ per year,][]{Dougherty2010AA...511A..58D,Fenech2018AA...617A.137F,Andrews2019AA...632A..38A}. These short-lived and episodic mass-loss events appear to be necessary to explain the diversity of evolved massive stars in Westerlund~1. \par

Westerlund~1 is rich in binary systems. A high binary fraction has been identified in massive stars through spectroscopic \citep{Ritchie2022AA...660A..89R}, radio \citep{Dougherty2010AA...511A..58D}, infrared \citep{Crowther2006MNRAS.372.1407C}, and X-ray \citep{Skinner2006ApJ...639L..35S,Clark2008AA.477.147C,Clark2019AA...623A..83C} observations. For instance, the WR population of Westerlund~1 has an estimated binary fraction of at least 70\% \citep{Crowther2006MNRAS.372.1407C,Clark2008AA.477.147C}. Isolated stars are primarily found among the mid-B to F hypergiants, with the exception of the LBV star W243, whose binarity is supported by interferometric \citep{Clark2019AA...623A..83C}, X-ray \citep{Mahy2022AA...657A...4M}, and spectroscopic \citep{Ritchie2009AA...507.1585R} observations.\par

This harsh environment is expected to have effects on the star formation process, the evolution and dispersal of protoplanetary disks, and the formation and early evolution of planets and their atmospheres. While no studies to date have been able to identify the population of protoplanetary disks in Westerlund~1 and explore the feedback provided by the starburst environment on their evolution and dispersal, several authors have attempted to quantify the cluster's initial mass function (IMF) to investigate possible deviations from the universal law. An IMF consistent with the \citet{Salpeter1955} law has been found by \citet{Brandner2008AA.478.137B} in the 3.4--27$\,$M$_\odot$ range, by \citet{Gennaro2011} extrapolated in the 0.5--120$\,$M$_\odot$ range, and \citet{Andersen2017AA.602A.22A} down to 0.15$\,$M$_\odot$ in the outer cluster, while a shallower IMF slope was found by \citet{Lim2013AJ....145...46L}, integrated in the 0.08-85$\,$M$_\odot$ range.

\subsection{Previous X-ray observations}
X-ray observations of young clusters provide valuable diagnostics for selecting pre-main-sequence (PMS) stars independently of the presence of circumstellar disks \citep[e.g.,][]{Montmerle1996}, down to low stellar masses \citep[e.g.,][]{GetmanFBG2005,Barrado2011AA...526A..21B}. Additionally, in a cluster rich in massive stars with a very compact configuration, X-ray observations can reveal a plethora of processes and physical mechanisms that play an important role in the evolution of massive stars \citep{Seward1979ApJ...234L..55S,Berghoefer1996AAS..118..481B}. It is also worth mentioning that only the \emph{Chandra} X-Ray Observatory \citep{WeisskopfBCG2002} can currently provide the high spatial resolution required to resolve individual X-ray faint sources in a crowded cluster like Westerlund~1. Given the designs of future X-ray missions currently in development, such observations will likely be challenging for quite some time after the \emph{Chandra} era. \par

Both \emph{Chandra} and XMM have been used in the past to observe Westerlund~1. The initial observations performed with Chandra reached a depth of approximately 58 ksec (P.I. Skinner) and resolved numerous X-ray sources \citep{Skinner2006ApJ...639L..35S,Muno2006ApJ.636L.41M,Clark2008AA.477.147C}. \citet{Skinner2006ApJ...639L..35S} focused on the WR stars and their spectral properties, detecting 12 out of 24 known stars and finding strong evidence for the existence of very hot plasma in the circumstellar environment in the two brightest objects (W72/A and WRB), strongly suggesting the presence of a colliding winds in these binary systems; \citet{Muno2006ApJ.636L.41M} studied the diffuse X-ray emission and its dominating hard spectral component, which was later confirmed by \citet{Kavanagh2011NewA.16.461K} from 48$\,$ksec XMM/Newton observations. These authors detected a strong Fe 6.7$\,$keV line in the diffuse emission spectrum, indicating its thermal nature. \citet{Clark2008AA.477.147C} found that 46 known high-mass members of Westerlund~1 were detected in X-rays, and they supposed that the remaining $\sim$60 X-ray sources detected in these images are likely PMS stars with masses $\leq$1.5$\,$M$_{\odot}$.\par

Since its discovery by \citet{Muno2006ApJ.636L.41M}, the magnetar CXO J164710.2-455216 (CXOU J16) in Westerlund 1 — the brightest X-ray source in the cluster — has garnered significant attention.
Dedicated observations using XMM-Newton and \emph{Chandra}/ACIS-S have accumulated a total exposure of 273.14$\,$ksec (P.I.s Israel, Muno and Schartel) and 94.65$\,$ksec (P.I.s Israel and Rea), respectively.
A typical property of this class of pulsars is their frequent bursts and recurrent outbursts. In fact, three distinct outbursts from CXOU J16 have been observed in the past 17 years \citep{Borghese2019MNRAS.484.2931B}. The first one occurred in September 2006 and was triggered by a short burst that released an energy of approximately $10^{39}\,$erg in the 15-150 keV band \citep{Krimm2006GCN..5581....1K}. It was followed  by a second outburst in September 2011 \citep{Israel2011ATel.3653....1I}, during which the pulsar exhibited a peculiar behavior: The pulse profile evolved from a single peak in the pre-outburst phase to an energy-dependent tri-peaked profile post-outburst. The overall spectrum evolved from a single blackbody to a more complex shape that was well modeled by including an additional hotter blackbody component. The most recent outburst was again triggered by a short burst detected by the \emph{Swift} Burst Alert Telescope in May 2017 \citep{Dai2017GCN.21095....1D}. During these intense outburst activities, the magnetic field strength was estimated to range from 7$\times$10$^{13}\,$G (a value typical for low-field magnetars, \citealp{PernaPons2011ApJ...727L..51P}) to $\sim$10$^{14}\,$G \citep{An2013ApJ...763...82A,Israel2007ApJ.664.448I}.  \par

\begin{table*}
\caption{Pre-EWOCS observations of Westerlund~1.}             
\label{tab_preewocsobs}      
\centering          
\begin{tabular}{|c|c|c|c|c|c|c|c|}    
\hline       
Obs.ID. & Instrument & Exposure & Roll Angle & RA   & Dec   & Date & P.I. \\ 
        &            & ksec     & degrees    &J2000 & J2000 &            &      \\
\hline                    
5411  &	ACIS-S & 38.47 & 326 & 16:47:05.40  &  -45:50:36.70 & 2005-06-18 & Skinner  \\
6283  &	ACIS-S &	18.81 & 25  & 16:47:05.40  &  -45:50:36.70 & 2005-05-22 & Skinner  \\
14360 &	ACIS-S &	19.06 & 242 & 16:47:10.20  &  -45:52:16.90 & 2011-10-23 & Israel   \\
19135 &	ACIS-S &	9.13  & 22  & 16:47:10.20  &  -45:52:17.00 & 2017-05-25 & Rea      \\
19136 &	ACIS-S &	13.67 & 331 & 16:47:10.20  &  -45:52:17.00 & 2017-06-16 & Rea      \\
19137 &	ACIS-S &	18.2  & 295 & 16:47:10.20  &  -45:52:17.00 & 2017-07-10 & Rea      \\
19138 &	ACIS-S &	18.2  & 86  & 16:47:10.20  &  -45:52:17.00 & 2018-02-24 & Rea      \\
20976 &	ACIS-S &	16.39 & 86  & 16:47:10.20  &  -45:52:17.00 & 2018-02-25 & Rea      \\
\hline                  
\end{tabular}
\end{table*}

\subsection{The EWOCS project}

The pre-EWOCS \emph{Chandra} observations of Westerlund~1 have been analyzed by \citet{Townsley2018ApJS.235.43T} in the framework of the Second Installment of the Massive Star-forming Regions Omnibus X-ray Catalog (MOXC2), identifying 1721 X-ray sources. This work has confirmed that Westerlund~1 is rich in X-ray bright sources, even though its low-mass stellar content remained undetected in the pre-EWOCS observations. According to these authors, the X-ray luminosity limit in the broad band where half of the brighter population is detected was log(Lx)=30.69, with Lx in erg/s, corresponding to a 1.5$\,$M$_{\odot}$ star\footnote{As stated by \citet{Townsley2018ApJS.235.43T}, the corresponding X-ray flux has been calculated using PIMMS6 assuming a limit of five-counts detection on-axis, for a source with an APEC thermal plasma with kT=2.7$\,$keV and abundance 0.4$\times$Z$_{\odot}$, which are typical values for a PMS star \citep{PreibischKFF2005}.}.\par

The need to unveil the low-mass population of Westerlund~1 has motivated the 1 Msec observation of Westerlund~1 (P.I. Guarcello) with the \emph{Chandra} Advanced CCD Imaging Spectrometer \citep[ACIS-I,][]{GarmireBFN2003}, which, together with a 18.9$\,$hours Cycle 1 JWST/MIRI and NIRCam observation (program ID 1905, P.I. Guarcello) and a 48 ksec NICER observation (P.I. Borghese) of CXOU~J16, constitutes the set of new observations of the EWOCS project\footnote{https://Westerlund1survey.wordpress.com/}. The main objective of EWOCS is to use Westerlund~1 and 2 as a test cases for understanding how star and planet formation, early stellar evolution, and the production of compact objects occur in a starburst environment. Specifically, the project aims to achieve the following objectives:
\begin{itemize}
\item Unveil the low-mass stellar population of Westerlund~1 and 2, both in their core and halo. X-ray observations are expected to be critically important for selecting cluster members in the halo, where contamination from background and foreground sources could affect membership determination based on photometric data. 
\item Determine the actual stellar content of the clusters, down to the low-mass regime, mainly thanks to the JWST observations; calculate their IMF down to the brown dwarf regime, and understand whether the starburst environment impact the formation of low-mass and very-low mass stars.
\item Study the clusters properties, particularly age, age spread, morphology and dynamics. The project aims to understand whether the clusters formed in a single burst of star formation or through a process spanning several million years, as well as how and if they will disperse. 
\item Identify the disk-bearing population of the clusters, mainly though the JWST observations. Combining this with the detection of disk-less stars from the \emph{Chandra}/ACIS-I observations and modeling of disks dispersal, we will finally assess how disks evolve and how planet formation proceeds in a starburst environment. 
\item If planets can form, understand how they evolve while immersed in such an environment characterized by high local fluxes of UV and X-ray radiation and relativistic particles.
\item Study how binarity and mass-loss affect the evolution of the massive stars in the clusters, and how their initial mass is mapped into the type of compact objects formed at the end of their evolution. 
\item Determine whether binarity across stellar masses is different in a starburst environment. 
\item Study for the first time the status of CXOU~J16 far from bursts, which will allow us to estimate the intrinsic properties of the pulsar. 
\item Search for the expected population of compact objects that have been suggested to exist in Westerlund~1, since, under specific assumptions, up to $\sim$65 core-collapse supernovae could have already occurred in the cluster \citep{Muno2006ApJ.636L.41M,Brandner2008AA.478.137B}. Besides, Westerlund~1 is one of the few known star clusters meeting the properties required for the formation of intermediate mass black holes from runaway coalescence \citep{PortegiesZwart2004Natur.428.724P}. As estimated by \citet{Clark2008AA.477.147C}, such objects, if present and if currently accreting mass, should be observable with a very deep \emph{Chandra} observation.
\item Understanding how stellar winds from massive evolved stars can affect the ISM to produce diffuse X-ray emission, whether this hot gas could affect star formation throughout the region, and whether we can prove ongoing accumulation of polluted material in the cluster core. 
\end{itemize}

Figure~\ref{fig_evtcontours} shows the contours of the pre-EWOCS and EWOCS observations of Westerlund~1 and CXOU~J16, plotted over the combined \emph{Chandra} event file and a $K_S$ band image of the cluster and the surrounding area obtained with the FourStar infrared camera mounted on the Magellan 6.5$\,$m telescopes. \par

\section{\emph{Chandra} observations and data reduction}
\label{sec_data}

The EWOCS survey also includes eight pre-EWOCS observations performed with ACIS-S, two of which were pointed at Westerlund~1 and six at CXOU~J16. These observations were conducted between June 2005 and February 2018 (Table \ref{tab_preewocsobs}). Additionally, 36 EWOCS observations were carried out with the ACIS-I detector from June 2020 to August 2021. The aim point of each ACIS-I observation was adjusted based on the nominal roll angle, as indicated in Table \ref{tab_ewocsobs}. This adjustment was crucial to for avoiding gaps that cover the cluster core, the pulsar, or some of the brightest massive members, while ensuring the cluster remained within the inner arcminute. By adopting this design, we maximized the benefits of the subarcsecond spatial resolution and sensitivity in the central part of the ACIS-I detector to observe the cluster core, which is highly compact and crowded. \par

\begin{table*}
\caption{EWOCS observations}             
\label{tab_ewocsobs}      
\centering          
\begin{tabular}{|c|c|c|c|c|c|}     % 7 columns 
\hline       
Obs.ID. & Exposure & Roll Angle & RA   & Dec   & Date \\ 
        & ksec     & degrees    &J2000 & J2000 &           \\
\hline                    
22316 & 39.55  & 245  &  16:46:59.97   & -45:51:13.70  &   2020-10-04  \\
22317 & 24.75  & 272  &  16:47:00.55   & -45:51:29.59  &   2021-08-14  \\
22318 & 26.72  & 312  &  16:47:03.24   & -45:51:45.84  &   2020-06-25  \\
22319 & 46.45  & 243  &  16:46:59.97   & -45:51:13.70  &   2020-10-09  \\
22320 & 37.58  & 321  &  16:47:05.45   & -45:51:39.01  &   2020-06-20  \\
22321 & 37.58  & 1    &  16:47:04.93   & -45:51:14.41  &   2020-06-02  \\
22977 & 37.57  & 236  &  16:46:59.97   & -45:51:13.70  &   2020-10-22  \\
22978 & 24.75  & 340  &  16:47:05.45   & -45:51:39.01  &   2021-06-12  \\
22979 & 21.79  & 14   &  16:47:07.63   & -45:51:13.62  &   2021-05-28  \\
22980 & 24.75  & 331  &  16:47:05.45   & -45:51:39.01  &   2021-06-16  \\
22981 & 21.85  & 314  &  16:47:03.24   & -45:51:45.84  &   2021-06-24  \\
22982 & 16.85  & 335  &  16:47:05.45   & -45:51:39.01  &   2020-06-13  \\
22983 & 27.72  & 340  &  16:47:05.45   & -45:51:39.01  &   2021-06-09  \\
22984 & 22.61  & 303  &  16:47:03.24   & -45:51:45.84  &   2021-07-02  \\
22985 & 24.75  & 52   &  16:47:07.13   & -45:50:48.14  &   2021-05-01  \\
22986 & 17.67  & 335  &  16:47:05.45   & -45:51:39.01  &   2020-06-11  \\
22987 & 24.75  & 1    &  16:47:04.93   & -45:51:14.41  &   2021-06-04  \\
22988 & 17.85  & 280  &  16:47:02.20   & -45:51:31.62  &   2021-07-27  \\
22989 & 21.79  & 14   &  16:47:07.63   & -45:51:13.62  &   2021-05-27  \\
22990 & 24.75  & 288  &  16:47:01.97   & -45:51:40.91  &   2020-07-17  \\
23272 & 11.92  & 1    &  16:47:04.93   & -45:51:14.41  &   2020-06-03  \\
23279 & 29.69  & 335  &  16:47:05.45   & -45:51:39.01  &   2020-06-12  \\
23281 & 30.49  & 321  &  16:47:05.45   & -45:51:39.01  &   2020-06-21  \\
23287 & 34.61  & 312  &  16:47:03.24   & -45:51:45.84  &   2020-06-26  \\
23288 & 29.18  & 312  &  16:47:03.24   & -45:51:45.84  &   2020-06-26  \\
24827 & 24.75  & 269  &  16:47:00.55   & -45:51:14.83  &   2021-08-21  \\
24828 & 24.75  & 1    &  16:47:04.93   & -45:51:14.41  &   2021-06-04  \\
25051 & 31.66  & 14   &  16:47:07.63   & -45:51:13.62  &   2021-05-28  \\
25055 & 29.68  & 1    &  16:47:04.93   & -45:51:14.41  &   2021-06-05  \\
25057 & 25.25  & 340  &  16:47:05.45   & -45:51:39.01  &   2021-06-13  \\
25058 & 27.22  & 333  &  16:47:05.45   & -45:51:39.01  &   2021-06-10  \\
25073 & 34.62  & 314  &  16:47:03.24   & -45:51:45.84  &   2021-06-25  \\
25096 & 18.14  & 280  &  16:47:02.20   & -45:51:31.62  &   2021-07-29  \\
25097 & 23.59  & 280  &  16:47:02.20   & -45:51:31.62  &   2021-07-30  \\
25098 & 25.43  & 280  &  16:47:02.20   & -45:51:31.62  &   2021-08-01  \\
25683 & 24.74  & 272  &  16:47:00.55   & -45:51:29.59  &   2021-08-15  \\
\hline                  
\end{tabular}
\end{table*}

\begin{figure*}[!h]
\centering
\includegraphics[width=0.45\textwidth]{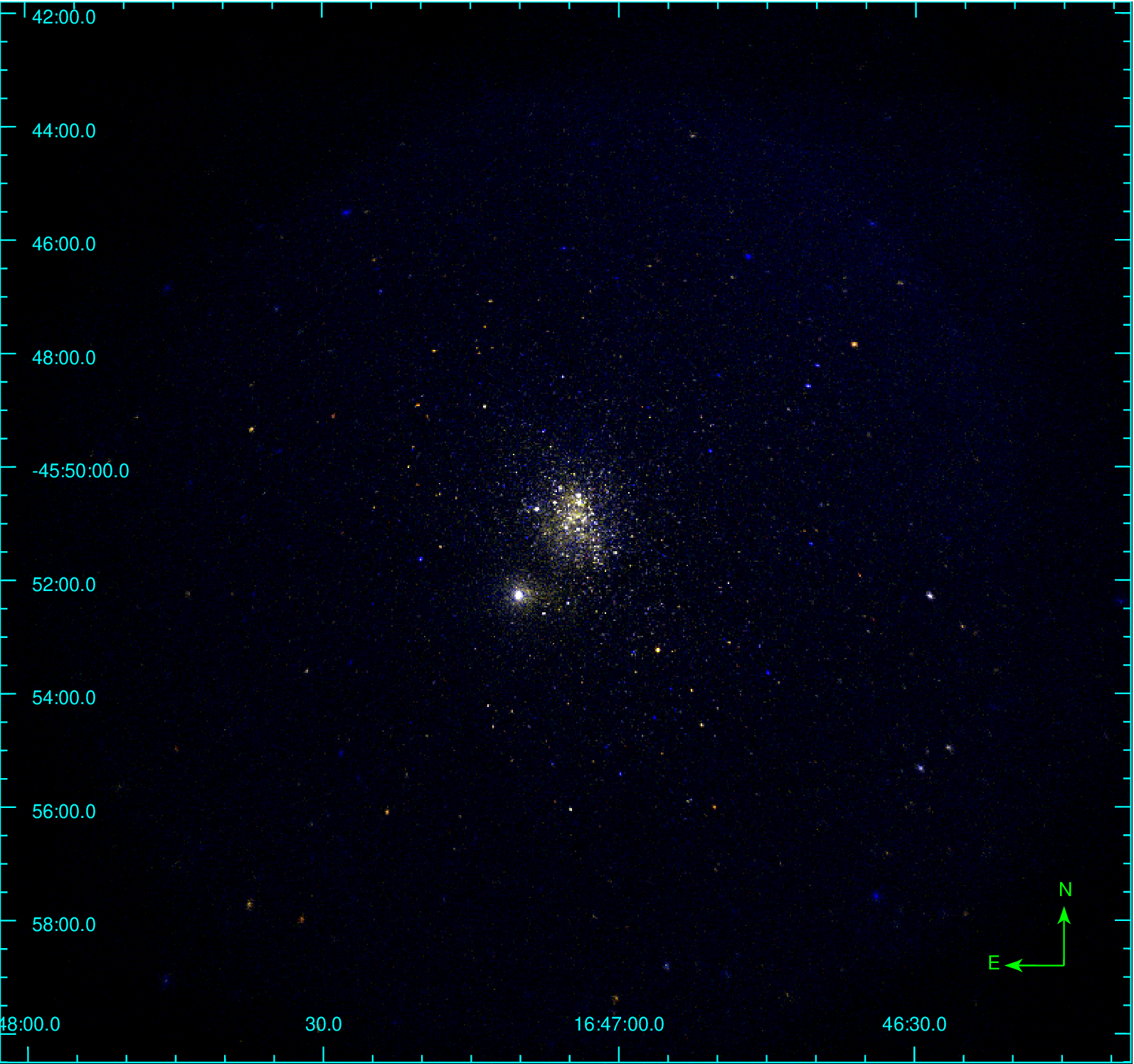}
\includegraphics[width=0.45\textwidth]{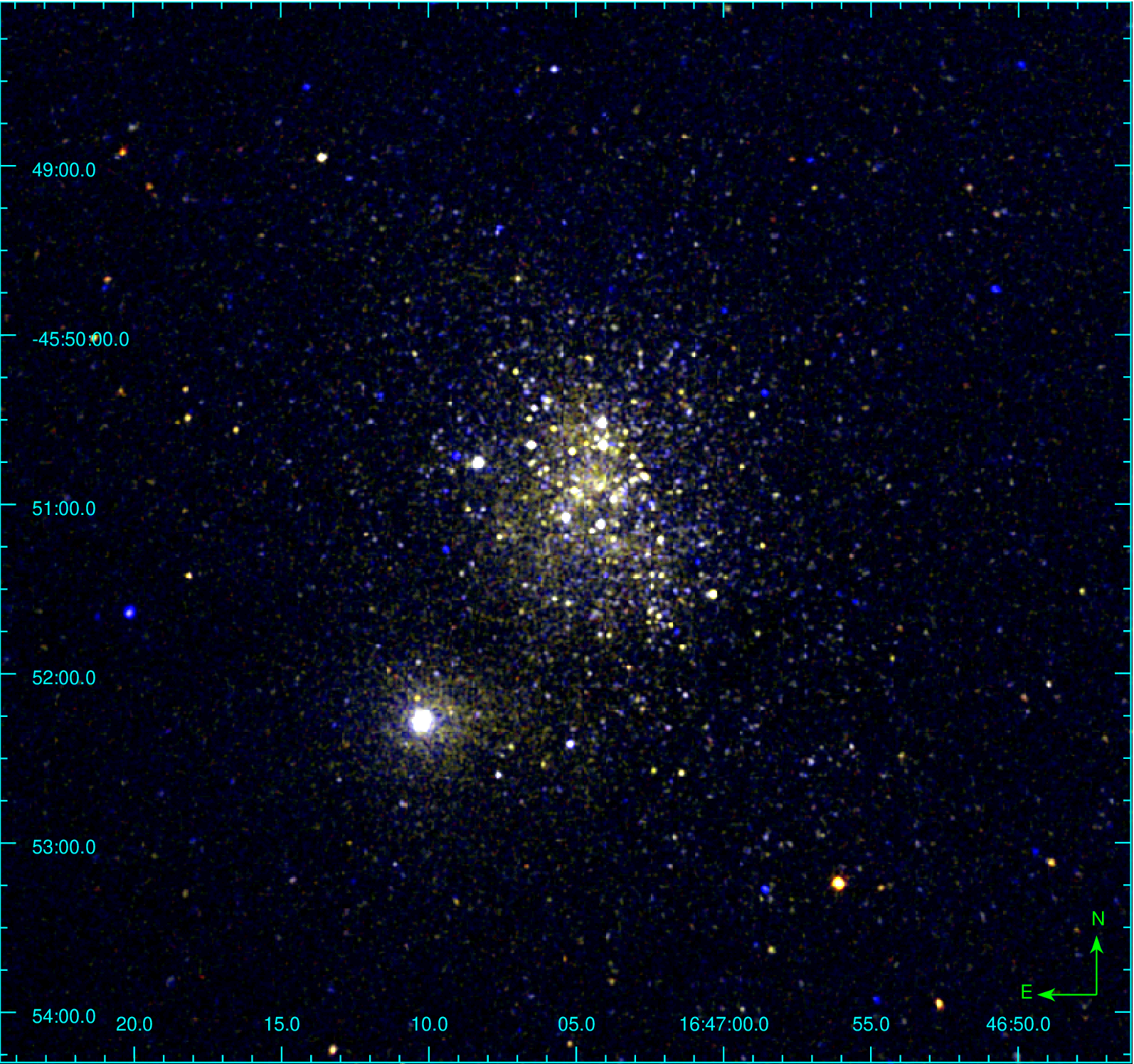}
\caption{RGB images of the whole ACIS-I field (left panel) and the central area (right panel) of the composite \emph{Chandra} images. Soft band (0.5-1.0$\,$keV) photons are marked in red, medium band (1.0-2.0$\,$keV) photons in green, and hard band (2-7.9$\,$keV) photons in blue. The brightest source in the southeast direction is CXOU~J16. The two images were smoothed adopting a Gaussian kernel with a radius of 2 pixels.}
\label{fig_rgbchandra}%
\end{figure*}

The total exposure for the pre-EWOCS observations is 151.93$\,$ksec, while for the EWOCS observations it is 967.80$\,$ksec. The exposure times for the individual EWOCS observations range from 11.92$\,$ksec (Obs.ID 23272) to 39.55$\,$ksec (Obs.ID 22319), with a mean exposure of 26.33$\,$ksec. The EWOCS observations span over one year, providing a robust baseline for studying the X-ray variability of the brightest sources. All EWOCS observations were conducted using the ACIS-I detector in imaging mode, utilizing all four chips (I0--I3). The observations were performed in the VERY FAINT mode, which employs telemetry in 5$\times$5 pixel event islands for improved background suppression\footnote{http://cxc.harvard.edu/cal/Acis/Cal\_prods/vfbkgrnd/index.html}. When combined with the pre-EWOCS observations, the total time baseline exceeds 16 years, which is particularly valuable for studying certain sources in the cluster, such as the magnetar. Figure \ref{fig_rgbchandra} displays a composite RGB ACIS image of Westerlund~1, where colors represent different photon energies (red: soft band, green: medium band, blue: hard band). The image shows both the entire EWOCS field and a central region of approximately $\sim$3$^{\prime}$. In the right panel, it is evident that the source density is high in the cluster core and it reveals that the majority of faint sources are predominantly hard, likely due to high absorption or since they have been observed during periods of intense magnetic activity such as flares.
The list of Chandra datasets used in this paper, and obtained by the Chandra X-ray Observatory, are contained in the Chandra Data Collection (CDC) 153~\href{https://doi.org/10.25574/cdc.153}{https://doi.org/10.25574/cdc.153}

\subsection{Data reduction}
The \emph{Chandra} observations were analyzed using the "pre-\emph{ACIS} Extract workflow" procedure outlined in \citet{Townsley2003ApJ...593..874T} and \citet{BroosTFG2010}. This procedure utilizes various tools integrated within the \emph{Chandra} Interactive Analysis of Observations (CIAO) software \citep{Fruscione2006SPIE.6270E.1VF}. We employed versions 4.13 of CIAO along with the CALDB 4.9.5 calibration files. \par

The L1-to-L2 processing flow aims to generate calibrated \emph{Chandra} event files from the L1 products provided by the \emph{Chandra} X-Ray Center (CXC). It includes event energy calibration, refinement of event positions, and correction for contamination caused by bad pixels and cosmic-ray afterglow. This workflow utilizes a less aggressive bad pixel table compared to the one produced by CIAO and it incorporates the $clean55$ algorithm for background reduction. Additionally, cosmic-ray afterglows are removed, and the source point spread function (PSF) is improved by disabling the random $\pm$0.25 pixel randomization. The standard grade filter is applied to events, retaining only $ASCA$ grades 0, 2, 3, 4, and 6. However, events are not filtered for the standard \emph{status=0} requirement, which may result in the exclusion of a significant number of reliable events.\par

Afterglows, which are groups of events appearing at the same location in consecutive CCD frames, can often be mistaken for faint sources. To address this, the CIAO tool \emph{acis\_detect\_afterglow} is typically employed to remove afterglows. This tool applies a relatively aggressive cleaning approach, eliminating several false positives. Another tool, \emph{acis\_run\_hotpix}, is less aggressive but it may fail to detect afterglow series with fewer than ten counts. In this L1-to-L2 procedure, a bifurcated workflow is adopted, where we applied an aggressive cleaning to the files used for source detection and validation, and a less aggressive cleaning for the files used in spectral analysis. \par

The background light curves were examined to identify and exclude intervals with intense and fluctuating background. This correction was required only for Obs.ID 5411, as the background remained relatively stable throughout the other observations. \par

The astrometry of the event files was corrected in three steps. In the first step, we addressed the offset of each Obs.ID relative to Obs.ID 22319, which is the deepest observation. We utilized \emph{Wavdetect} to identify the brightest sources in each observation and cross-matched their positions with those detected in the Obs.ID 22319 image. Subsequently, we employed the CIAO tools \emph{wcs\_match} and \emph{wcs\_update} to update the astrometry for each observation. In the second step, which was part of the L1-to-L2 workflow, we corrected the astrometry of each event file using the \emph{Gaia} Third Data Release \citep[DR3;][]{GaiaCollaboration2023A&A...674A...1G} astrometric system. This process was repeated as the third step, but using the brightest sources from the final list of validated sources (Sect. \ref{sec_finalcatalog}). \par  

\begin{figure}[!h]
\centering
\includegraphics[width=0.48\textwidth]{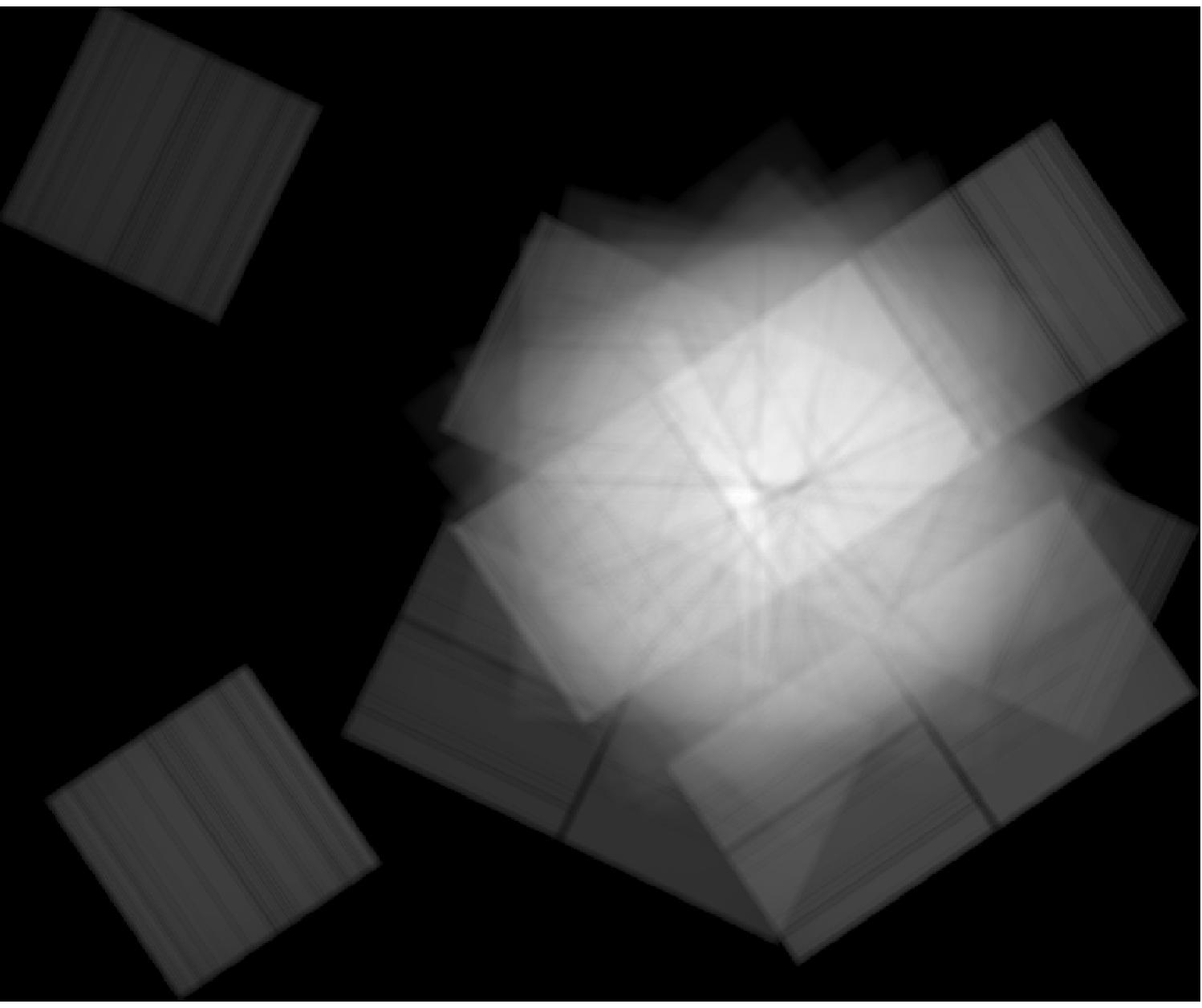}
\caption{Combined exposure map in the broad band.}
\label{fig_emap}%
\end{figure}

Exposure maps were calculated using the standard CIAO tools implemented in the pre-\emph{ACIS} Extract workflow for each observation in the broad (0.5--7.9$\,$keV), soft (0.5--1.0$\,$keV), medium (1.0--2.0$\,$keV), hard (2.0--7.9$\,$keV), and very hard (4.0--7.9$\,$keV) bands, and subsequently combined. The resulting combined exposure map in the broad band is displayed in Fig. \ref{fig_emap}, revealing a deep and nearly uniform exposure in the central region. This region is sufficiently large to encompass both the core of Westerlund~1 and a portion of the expected halo of the cluster \citep[as recently discovered, extended haloes are typically associated with stellar clusters;][]{Meingast2021AA...645A..84M,Prisinzano2022AA...664A.175P}.

\section{Source detection}
\label{sec_detection}

\begin{figure*}[!h]
\centering
\includegraphics[width=0.98\textwidth]{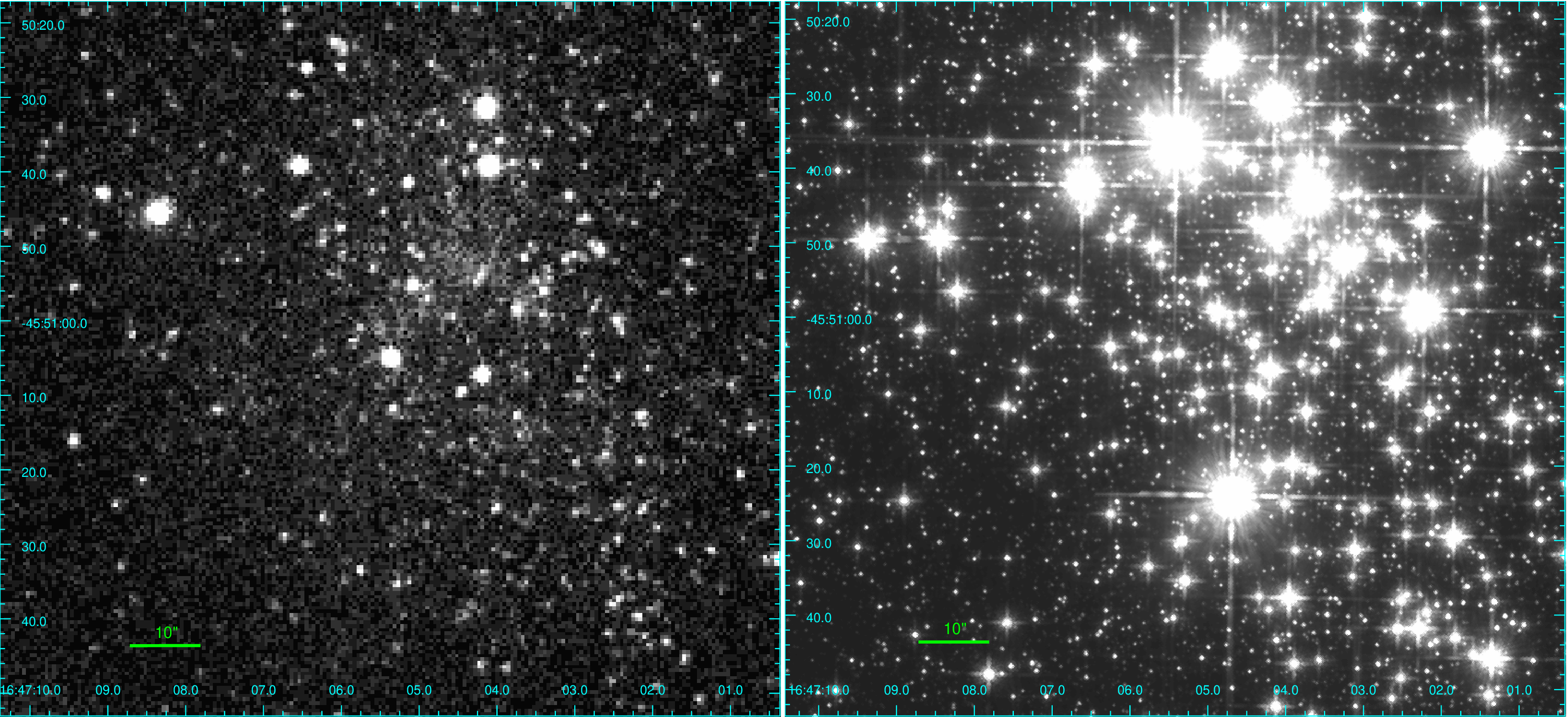}
\caption{Inner region of Westerlund~1 observed with ACIS (left panel) in the broad band and with HST (right panel) using the F160W filter. In the ACIS image, source confusion and a high background intensity dominate the cluster core.}
\label{fig_centerHST}%
\end{figure*}

The strategy we employed for source detection aims to maximize the depth of the EWOCS catalog, even in the core of Westerlund~1. This region presents challenges due to source confusion and a bright, irregular background, making the detection of faint sources a complex task (see Figure \ref{fig_centerHST}).

Source detection is implemented using four different methods:
\begin{itemize}

\item The wavelet-based algorithm \emph{PWDetect} \citep{Damiani1997ApJ.483.350D} is applied in the broad, soft, medium, hard, and very hard energy bands. The detection threshold we adopted roughly corresponds to 50 spurious sources. We excluded the outermost regions where the selection resulted in a very large number of false positives, resulting in 2306 detected sources.

\item The wavelet-based algorithm \emph{Wavdetect} \citep{Freeman2002ApJS.138.185F} is applied to images in the broad, soft, medium, hard, and very hard energy bands. We set the \emph{sigmathreshold} parameter equal to 10$^{-4}$ and used only two small detection scales, resulting in 2509 detected sources.

\item The maximum likelihood reconstruction method developed by \citet{Townsley2006AJ....131.2140T} is applied in the broad, soft, hard, and very hard energy bands. This algorithm operates over small tiles across the observed field, making it more sensitive to the spatial variation in the PSF and background and thus more capable of detecting faint sources in crowded fields \citep{BroosTFG2010}. The reconstructed image is first calculated using the Lucy-Richardson algorithm \citep{Lucy1974AJ.....79..745L}, and then searched for peaks that identify the positions of point sources. This method resulted in 7585 detected sources.

\item A time-resolved deployment of \emph{PWDetect}, described in more details in the following, is performed over segments of 10$\,$ksec of the observations. This method is aimed at detecting faint and variable sources that may only be significant during specific short time segments in which they were detected. This method produced a list of 1147 detected sources.
\end{itemize}
    
\begin{figure*}[!h]
\centering
\includegraphics[width=0.7\textwidth]{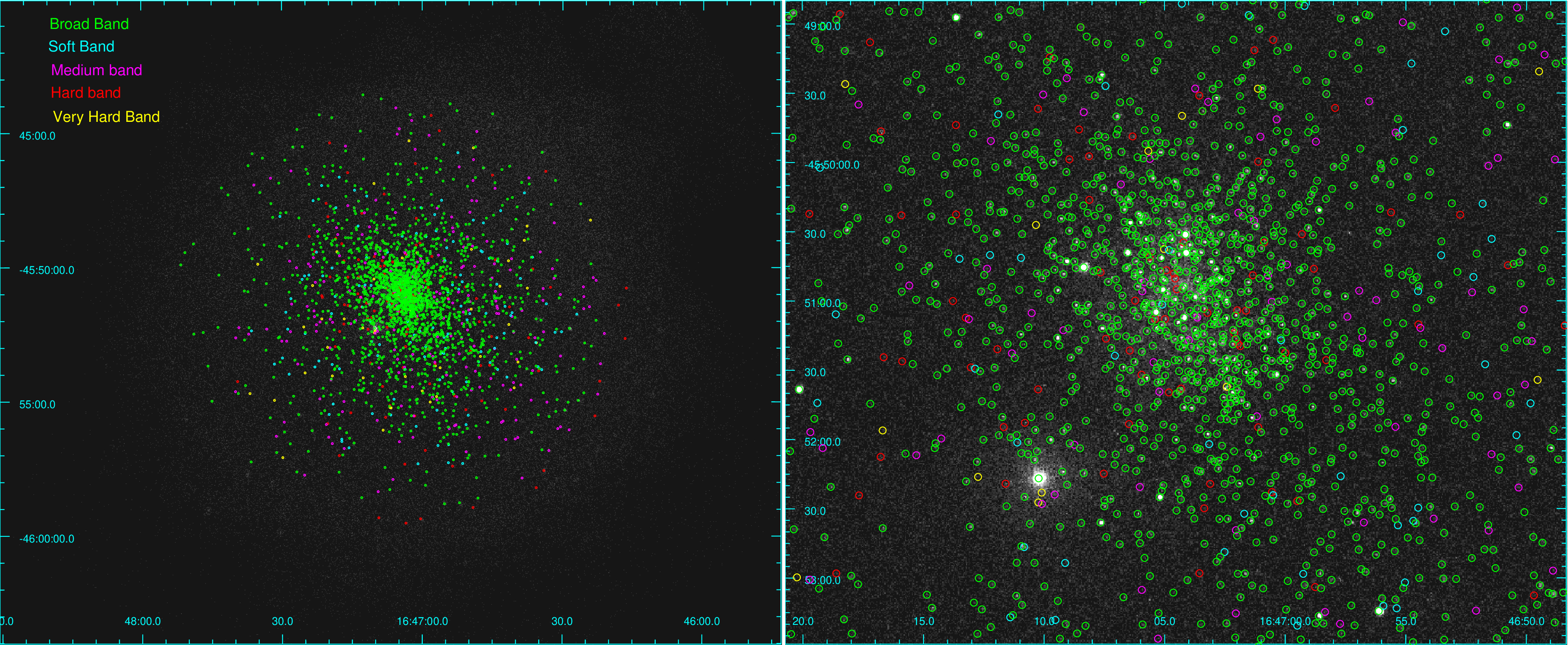}
\includegraphics[width=0.7\textwidth]{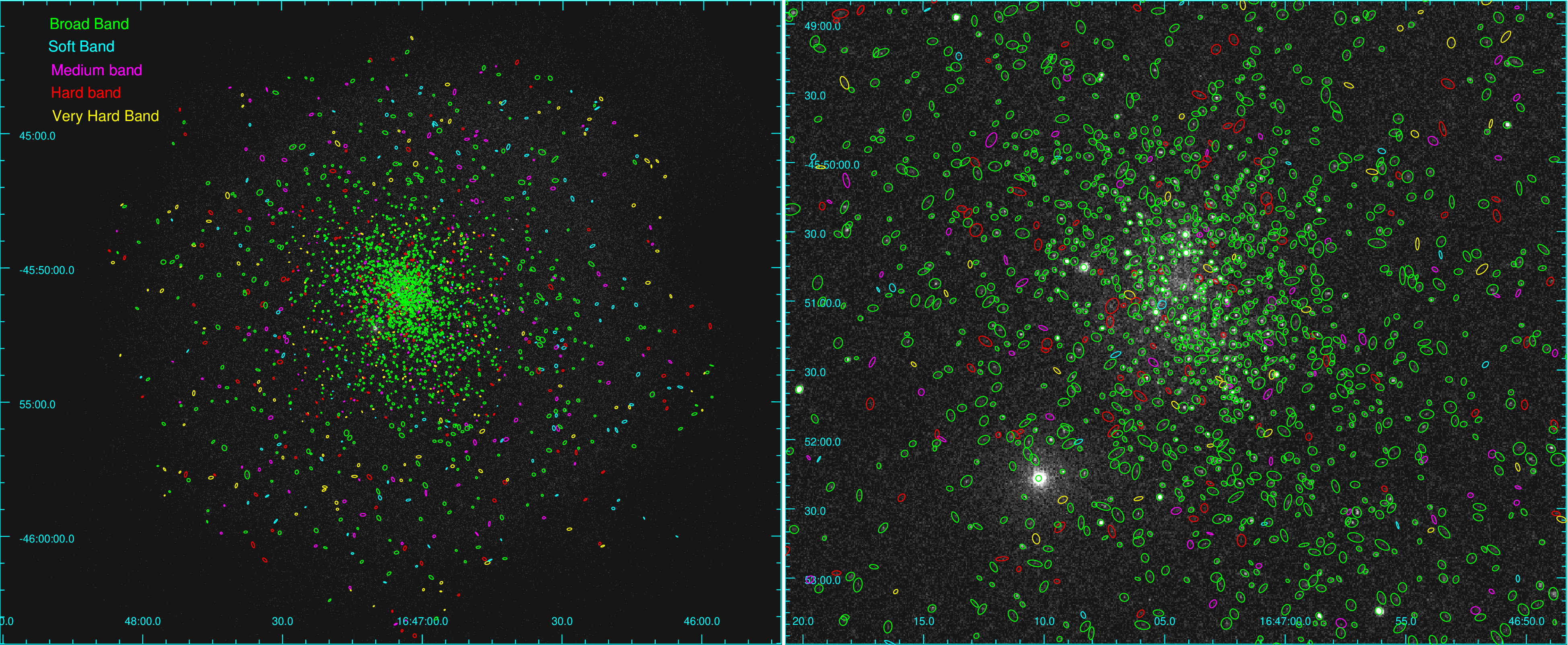}
\includegraphics[width=0.7\textwidth]{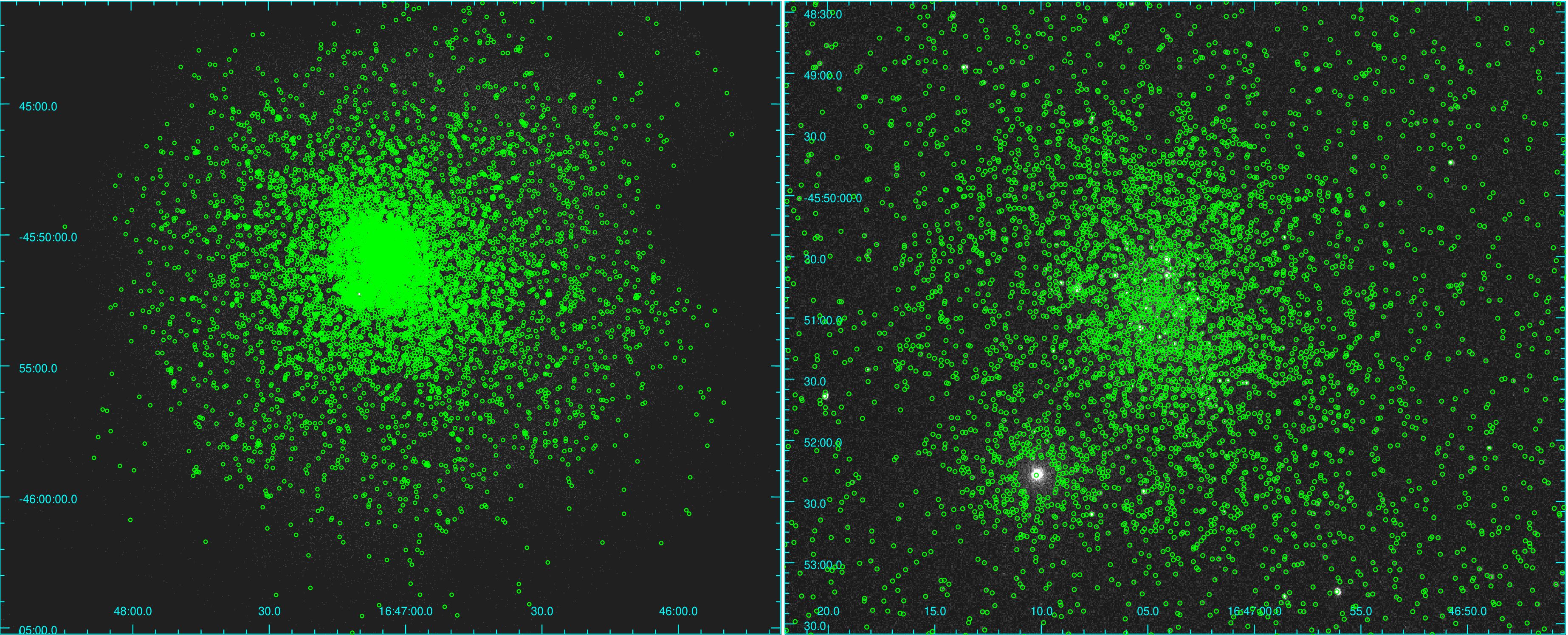}
\includegraphics[width=0.7\textwidth]{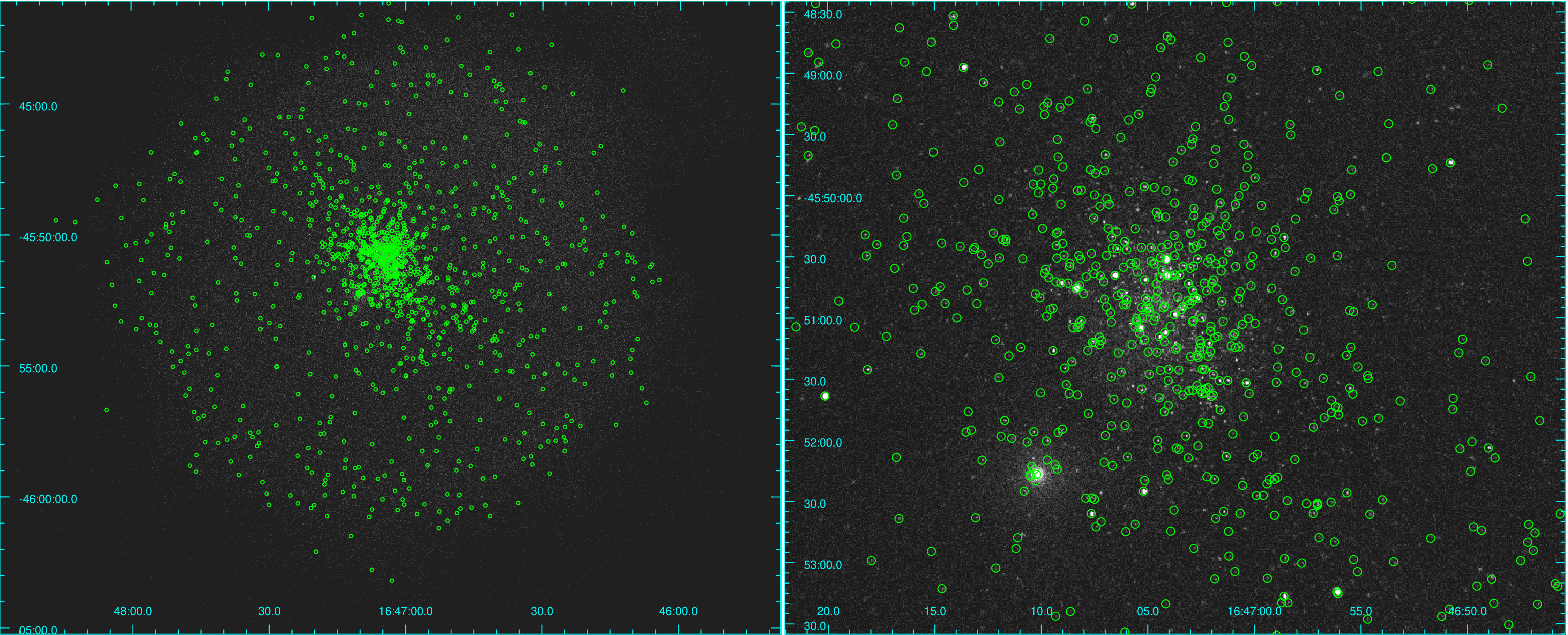}
\caption{Spatial distribution of candidate sources detected with the four methods (from the top: \emph{Pwdetect}, \emph{Wavdetect}, image reconstruction, and time-resolved \textit{Pwdetect}). The left panels show the whole ACIS field, those on the right the inner region. Different colors in the first and second rows mark sources detected at different energy bands.}
\label{fig_detmethods}%
\end{figure*}

Figure~\ref{fig_detmethods} shows a comparison of the spatial distribution of candidate sources detected using the four methods. In all cases, the cluster appears highly crowded, with the image reconstruction method being the only one capable of detecting a large number of sources in the central region of Westerlund~1, as expected. 

\subsection{Time-resolved PWDetect}

We devised a simple time-resolved detection method, tailored to faint transient sources such as magnetically flaring low-mass stars. These may remain undetected in the full dataset because of the high background, but may be detected in a shorter time slice that includes the transient emission, thanks to the enhanced source-counts/background contrast.

We started by considering 10$\,$ks time slices from each observation segment. Since exposure times are not multiple of 10$\,$ks, the exposure time of the last frame was forced to range between 7$\,$ks and 17$\,$ks. Moreover, in order to fully capture transients that would otherwise be split between two frames, we also considered intervals shifted in time by half a frame (5$\,$ks). 

The 44 EWOCS and pre-EWOCS observations were thus split into 194 frames: 106 were 10$\,$ks long, while the duration of the remaining ones are quite uniformly distributed between 5 and 16$\,$ks. For each frame, event lists (and exposure maps) were then extracted in the following five energy bands: 0.5-7.0$\,$keV, 4.0-7.0$\,$keV, 0.5-1.2$\,$keV, 1.2-2.0$\,$keV, and 2.0-7.0$\,$keV, resulting in a total of 970 event files.

We ran \emph{PWDetect} twice on each of these 970 event files, once to estimate the background level and thus evaluate the significance thresholds to obtain the desired number of spurious sources, and once more for the final detection. For the first run we adopted a low significance threshold, 4.9$\sigma$, so to detect as many sources as possible, but also resulting in several spurious faint sources. The number of background photons was then estimated by subtracting the detected source photons from the total; we then chose the final detection threshold so to yields, on average, 0.1 spurious sources per frame. This was derived from the appropriate significance versus background curve provided by \citet{Damiani1997ApJ.483.350D}. 

The final \emph{PWDetect} runs produced 944 lists of sources\footnote{For the observations in standard ACIS-I configuration, detection was performed only on the most on-axis CCD (CCD.ID=7). In 26 cases \emph{PWdetect} crashed or no sources were found. We did not investigate these cases further} for a total of 14178 sources, most of which are, obviously, repeated detections of the same source in multiple frames and/or bands. We started cleaning up this large sample by removing $\sim$1600 extended sources, many of which were unresolved detections of multiple point sources (extent parameter, as given by \emph{PWDetect}, larger than 2). We then screened for the remaining sources for possible cosmic rays afterglow events: for each source we extracted photons from a circle with radius twice the "detection scale" provided by \emph{PWDetect}. In 284 cases the arrival times of all extracted photons were in subsequent 3.14s-long readout frames, and the detection, a likely afterglow artifact was
discarded.

All detection lists were cross-identified and merged in a final source list using an iterative procedure: first we cross-identified and merged the first two catalogs. The resulting catalog was then merged with the third original catalog, and so on for all the 944 catalogs. Identifications were performed searching the close spatial coincidences with identification radii of each original detection taken as the 1$\sigma$ uncertainties as estimated by \emph{PWDetect} (rounded up to 0.5 arcsec if smaller). The coordinates and uncertainties or identification radii of merged sources were computed, at each step, as the uncertainty-weighted means of the coordinates/radii of cross-identified sources\footnote{Since most detections are not independent (they may share the same photons because of energy band or overlapping time frames), we computed weighted means only among values belonging to independent groups of positions. Within each group of dependent detections we chose the coordinates and radii of the source with the smallest positional uncertainty.} At the end of this process we are left with 1262 cross-matched sources.

Finally, we inspected all the final sources by eye, examining individual detections in the original event file, and the positions in the \emph{Hubble} Space Telescope (HST) H-band image (when available, in the field center) and images from the Digitized Sky Survey (DSS) and the Two Micron All-Sky Survey (2MASS). Some cross-identifications were adjusted and a number of "sources," which were not merged by the automatic process above, where merged as they clearly referred to the same star. The final list counts 1147 sources.

\subsection{The merged list of candidate sources}

At this stage, we generated a list of candidate sources that includes all the sources detected using the adopted methods. Contamination of this list by false positives is expected to be large, but we relied on the source-validation step, described in the next sections, to prune the catalog from these false and not significant sources. \par

The lists generated by the four detection methods were cross-matched by eye. In cases where it was difficult to confidently determine the presence of one or more nearby sources, we left those entries unmatched between the catalogs. Additionally, we included in the input list the following sources that were not detected by any of the aforementioned methods:

\begin{itemize}
\item 446 faint sources from the catalog presented by \citet{Townsley2018ApJS.235.43T}, which are likely not detected in EWOCS observations because of the intrinsic variability of young stars;   
\item 21 massive stars of Westerlund~1 from the list published by \citet{Clark2020AA...635A.187C};
\item 47 candidate sources added by eye corresponding to the positions of \emph{Gaia} sources in or nearby the cluster center. 
\end{itemize}

The final list of candidate sources, which was used as input for the source validation process in $ACIS$-Extract (AE), consists of 9420 sources.

\section{Source extraction, validation, and photometry}
\label{sec_extraction}

Source validation and photometry were performed using the AE software in IDL \citep{BroosTFG2010}\footnote{http://www.astro.psu.edu/xray/acis/acis\_analysis.html}, which has been successfully employed in previous X-ray surveys including the \emph{Chandra} Carina Complex Project \citep{TownsleyBCF2011}, the Massive Young Star-forming Complex Study in Infrared and X-Rays (MYStIX) survey \citep{Feigelson2013ApJS.209.26F}, the three Massive Star-forming Regions Omnibus X-ray Catalog (MOCX) data releases \citep{Townsley2014ApJS..213....1T, Townsley2018ApJS.235.43T, Townsley2019ApJS.244.28T}, the \emph{Chandra} Cygnus~OB2 Legacy Survey \citep{WrightDGA2014}, and the Star Formation In Nearby Clouds (SFiNCs) project \citep{Getman2017ApJS..229...28G}. AE enables the extraction and validation of sources across multiple observations, generating individual source spectra and light curves. It utilizes various data analysis software packages including CIAO, MARX \citep{Davis2012SPIE.8443E..1AD}, HEASoft\footnote{https://heasarc.gsfc.nasa.gov/lheasoft}, and the IDL Astronomy User's Library \citep{Landsman1993ASPC...52..246L}. \par

Following the guidelines provided by the authors and available on the AE website, we adopted a three-step procedure to compile the X-ray EWOCS catalog:

\begin{itemize}
\item Initially, sources were extracted and validated using a parameter defined by AE, which helps distinguish between genuine and false sources. This step was repeated iteratively until no more false sources were identified and removed (Sect. \ref{sect_sourcevalidation}).
\item Subsequently, source positions were updated, followed by another round of source validation process (Sect. \ref{sect_positions}).
\item Once the catalog reached a stable state, we performed the photometric procedure, to extract source events comprehensively and calculate the primary spectral and temporal properties for each source across multiple energy bands (Sect. \ref{sect_spectralextraction}).
\end{itemize}

\subsection{Source validation}
\label{sect_sourcevalidation}

The AE procedure assesses the local PSF  at the given position of each source and defines extraction regions based on the 1.5 keV local PSF, ensuring they do not overlap with neighboring sources. In the case of close pairs, the extraction region of the fainter source is progressively reduced to prevent overlap until it reaches 40\% of its original size. Once this threshold is reached, if the two extraction regions still overlap, AE further reduces the size of the brighter source until the regions no longer overlap. If overlap persists even when both extraction regions are reduced to 40\%, AE either discards the specific observation or automatically removes the fainter source.\par

The local background is determined within an optimized region surrounding the source. For isolated sources, this region is delimited by an inner radius, which is 1.1 times the radius encompassing 99\% of the PSF, and an outer radius large enough to collect at least 100 background events not associated with nearby sources. AE adjusts the size of the background-extraction region to ensure that Poissonian noise contributes no more than 3\% to the background uncertainty. However, in crowded regions, defining a region with 100 events may not be feasible. In such cases, AE employs a different calculation that incorporates the contribution from nearby bright sources and a model accounting for the spatial variation of the background.\par

Source validation relies on a parameter provided by AE called \emph{prob\_no\_source} (P$\rm_B$), which represents the probability that there is no real source at a given position. In our case, where multiple observations of a source are available, AE calculates P$\rm_B$ based on the extractions with the highest source significance. To differentiate between valid and spurious sources, we applied a threshold of P$\rm_B$=0.01, consistent with previous studies. Since the removal of not valid sources could potentially impact the extraction region and background of valid sources, the procedure is iterated until the catalog reaches convergence and no further spurious sources are detected.\par

\begin{figure}[!h]
\centering
\includegraphics[width=0.4\textwidth]{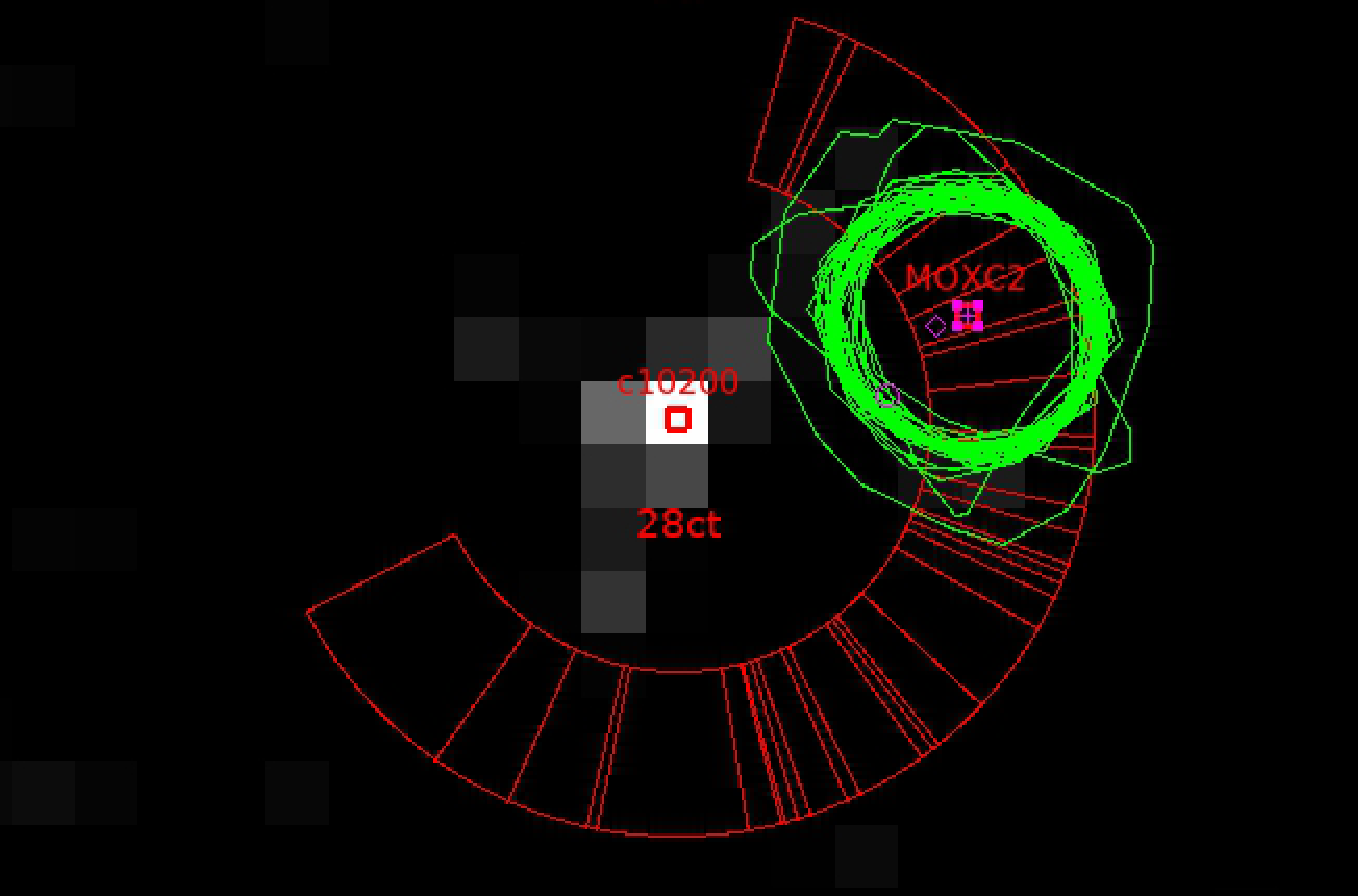}
\caption{Example of a source (label MOXC2) that was excluded as a potential product of a PSF hook near a brighter source (label c10200). The green contours outline the extraction regions of MOXC2 in all observations, while the red polygons indicate the locations where the PSF hook may appear in each observation, depending on the roll-angle.}
\label{fig_hook}%
\end{figure}

After the first iteration, we conducted a visual inspection of sources flagged by AE as potentially resulting from the hook-shaped feature of the PSF\footnote{http://cxc.harvard.edu/ciao/caveats/psf\_artifact.html}. This feature can account for up to 5\% of the source flux and its position is influenced by the roll-angle, making it distinguishable from the actual source only in a few cases where the real source is both on-axis and sufficiently bright. Figure~\ref{fig_hook} illustrates an example of a source (MOXC2) that was flagged by AE as a potential PSF hook and subsequently removed after visual inspection.

AE also identifies sources that are expected to suffer from significant pileup\footnote{http://cxc.harvard.edu/ciao/why/pileup\_intro.html} (which is the loss of information due to different incident photons registered as a unique event by the detector). In our case, the only source affected by piled source is the magnetar.\par

\subsection{Positions update and visual review}
\label{sect_positions}

For each source, AE calculates three different position estimates. The first estimate is obtained by taking the mean value of the positions of the events associated with the source (mean-data position). However, this estimate may be inaccurate for large off-axis angles and in cases where there are significant offsets between the true source position and the extraction region (which can happen when the PSF is asymmetric). To obtain a more accurate estimate in these cases, AE correlates the source PSF with the spatial distribution of extracted events (PSF position). This calculation takes into account the combination of several Obs.IDs by using the PSF calculated in each observation. Both of these estimates can be influenced by nearby sources. In crowded fields, a third estimate is provided by AE using the reconstructed image of the source's neighborhood. It identifies the position of the closest peak in the reconstructed image. According to AE's recommendations, the mean-data position is used for on-axis sources, the PSF position is used for off-axis sources, and the image reconstruction position is used for sources in crowded regions. The repositioning of sources was performed twice, with each step followed by a new sequence of iterations for source validation, as described in the previous section. \par

Before conducting the visual review of validated sources, the astrometry of both the X-ray sources and the main products file was corrected using the \emph{Gaia}/DR3 astrometric system. After this step, and when catalog stability was achieved again, we conducted a visual review of specific critical sources, including very faint sources that could affect the size of the extraction region of nearby bright sources and suspected afterglows. The decisions made during the visual review were guided also by the presence of high-probability optical and/or infrared counterparts. After the visual review, a new round of source validation was performed.\par

\subsection{Spectral extraction}
\label{sect_spectralextraction}

After 21 iterations of the source validation process, the catalog reached stability, with a total of 5963 validated X-ray sources. The final step involved the extraction of X-ray events and the estimation of X-ray properties in 17 energy bands, merging all available observations in a consistent manner. In addition, AE generates light curves and spectra for each source, although these will not be discussed in this paper. AE performs this calculation by excluding observations where the sources are observed off-axis to improve the overall signal-to-noise ratio. However, in our case, this correction was not necessary due to the design of our survey. The calculated quantities include source counts, net counts, photon flux in photon/cm$^2$/s, and the quartiles of photons energy. \par%Additionally, the resulting K-S probability is provided to assess whether the photon arrival times can be described by a constant flux. \par

\begin{figure*}[!ht]
\centering
\includegraphics[width=0.4\textwidth]{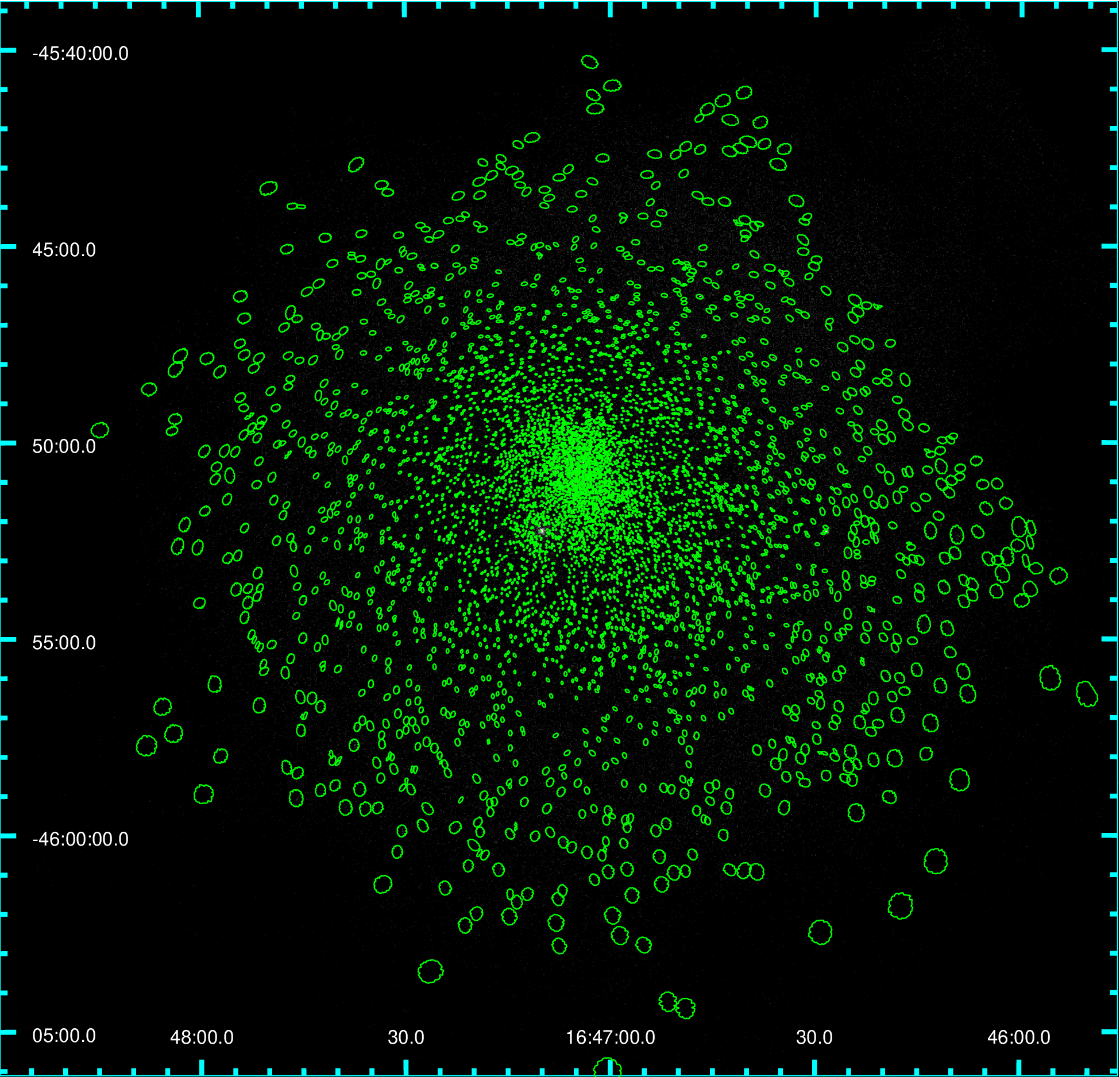}
\includegraphics[width=0.4\textwidth]{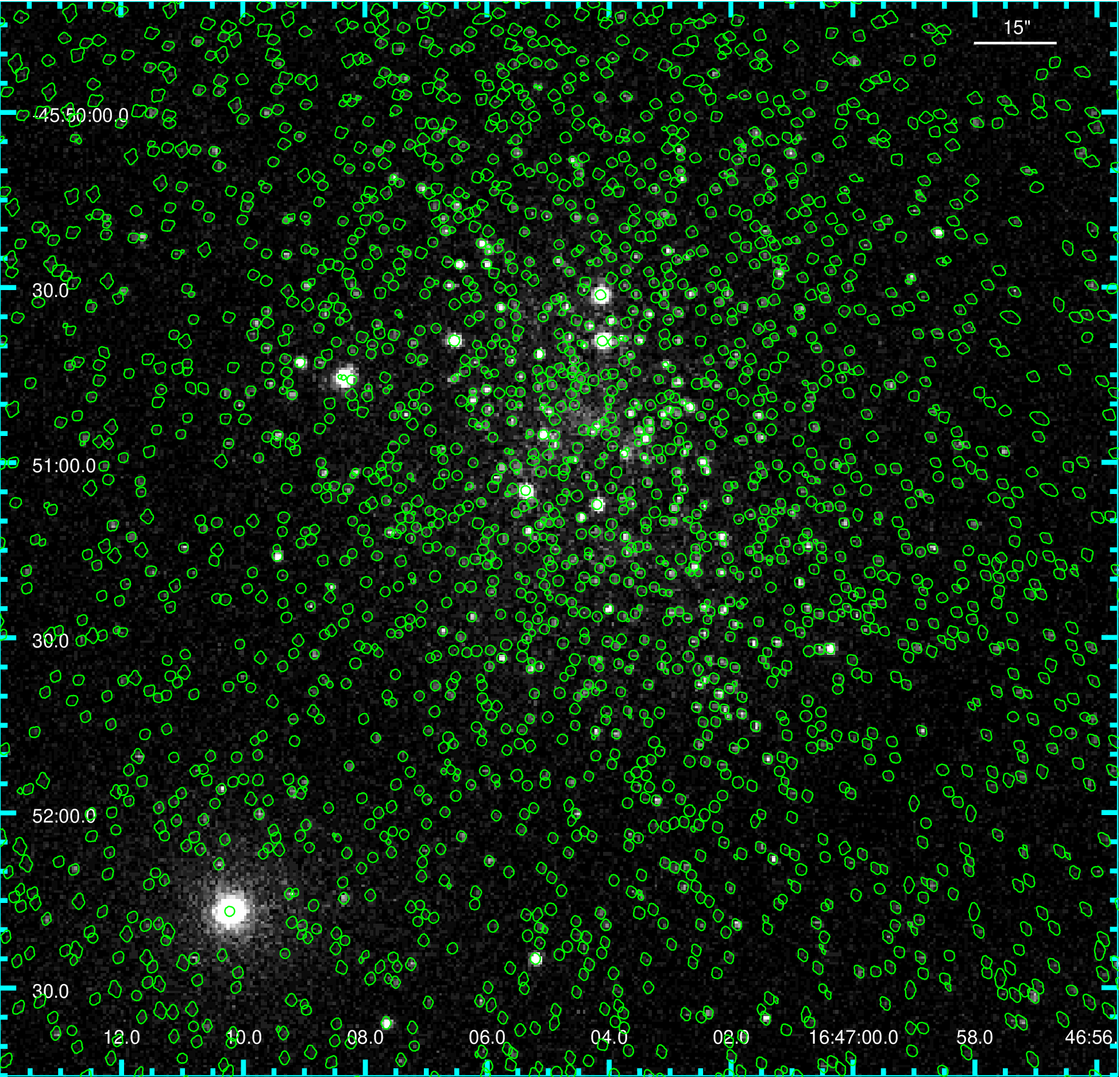}
\caption{Extraction regions of the validated sources across the entire merged ACIS image (left panel), and extraction regions in the central area of approximately $3^{\prime}$ in size (right panel).}%
\label{fig_validatedspadis}
\end{figure*}

\begin{figure}[!h]
\centering
\includegraphics[width=0.4\textwidth]{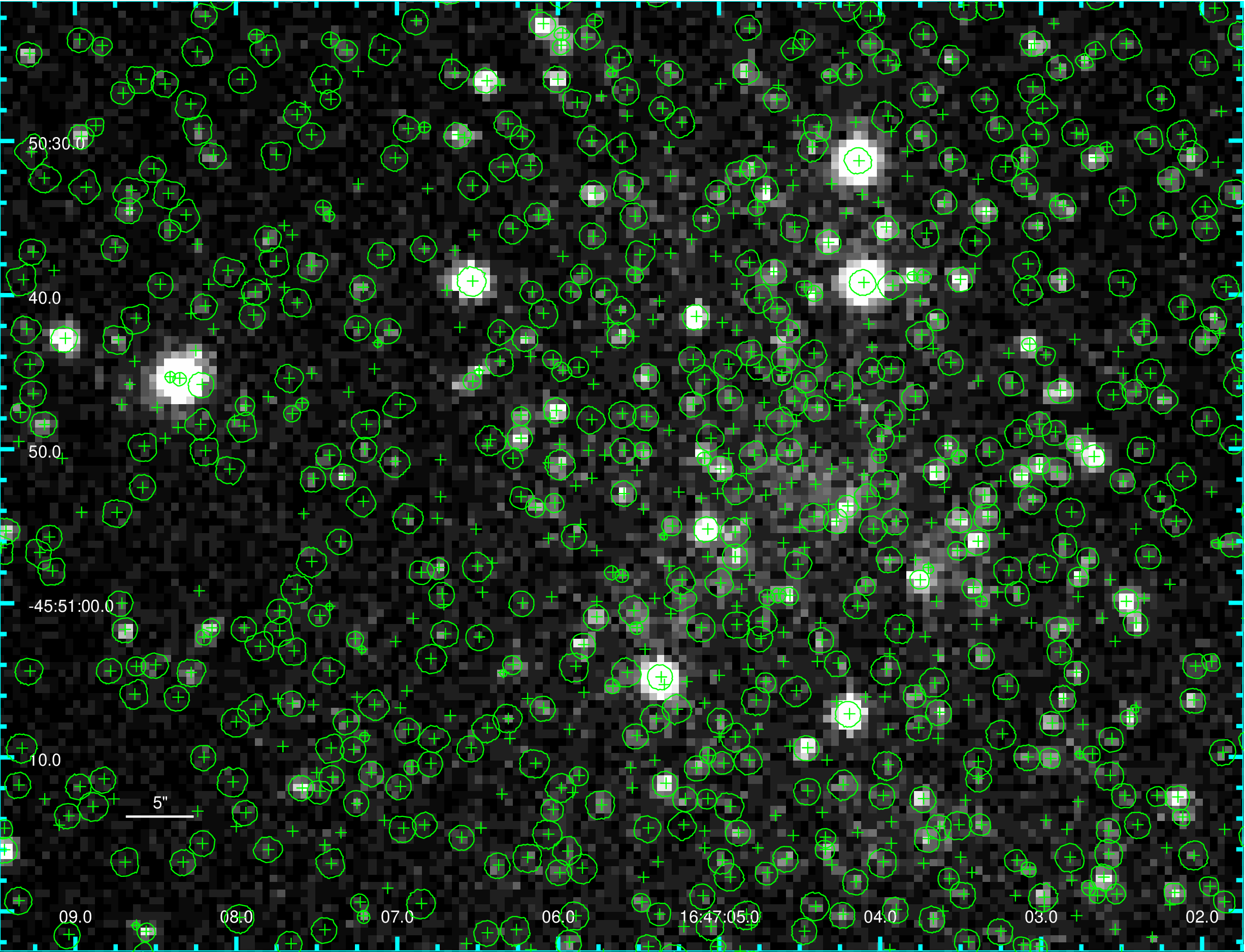}
\caption{Extraction regions of the validated sources and positions of the input candidate sources (crosses) within the central 1 arcmin region}
\label{fig_validatedandinput}%
\end{figure}

Figure~\ref{fig_validatedspadis} depicts the spatial distribution of the validated sources in the merged ACIS event files. In the left panel it is evident that there is a high concentration of validated sources toward the center of the cluster, as well as a significant number of sources surrounding the cluster core. This indicates that we have detected stars associated with the extended halo of Westerlund~1. This will be further investigated in upcoming papers of this series, which will focus on source classification and the identification of optical/infrared (OIR) counterparts. The right panel also highlights how this survey has pushed to the limits of \emph{Chandra} in resolving individual stars within such a densely populated stellar cluster with a bright and irregular background. In fact, in the actual core of the cluster, where the background is both intense and variable, a few tens of sources that were initially included as input to AE were subsequently discarded during the validation process (see Fig. \ref{fig_validatedandinput}). The limited number of validated sources in the central region can be attributed to the intense background, and it is likely that many of these discarded sources are indeed genuine X-ray sources. Although we did not attempt to recover these stars, in future papers of this series their candidate OIR counterparts will be analyzed in order to estimate the fraction of real sources that we have excluded.\par

\section{The final catalog}
\label{sec_finalcatalog}

It is informative to analyze the number of sources detected using the various methods we employed and assess how many have survived the pruning process. Table \ref{tab_ewocsrc} presents the total number of input sources for each detection method, as well as the fraction of these sources within 1$^{\prime\prime}$ and 3$^{\prime\prime}$ of a source in the final catalog (source positions changed during the pruning process and thus an exact position match was not possible). The image reconstruction method is the only one that experienced significant pruning of the input catalog, as it selects sources that are too faint according to the adopted P$\rm_B$ threshold. According to Table \ref{tab_ewocsrc}, and considering also that the number of sources in the input \emph{PWDetect} list of candidate sources more distant than 3$^{\prime\prime}$ from any source in the Image reconstruction input catalog is low, but not negligible (314, 688 for \emph{Wavdetect}), it is evident that in complex fields like these, deploying different detection methods is crucial for optimizing the number of detected sources.

\begin{table}
\caption{EWOCS sources and detection methods.}
\label{tab_ewocsrc}      
\centering          
\begin{tabular}{|c|c|c|c|}     % 7 columns 
\hline       
Detection method & Input N & within 1$^{\prime\prime}$ & within 3$^{\prime\prime}$\\ 
\hline                    
Image reconstruction                & 7585  & 0.29 &  0.30    \\
\emph{PWDetect}                     & 2306  & 0.87 &  0.92    \\
\emph{Wavdetect}                    & 2509  & 0.77 &  0.82    \\
Time resolved \emph{PWDetect}       & 1147  & 0.77 &  0.82    \\
Massive stars                       & 21    & 0.34 &  0.81    \\
\citet{Townsley2018ApJS.235.43T}    & 446   & 0.48 &  0.66    \\
Added by eye                        & 47    & 0.38 &  0.95    \\
\hline                  
\end{tabular}
\end{table}

\begin{figure}[!h]
\centering
\includegraphics[width=0.45\textwidth]{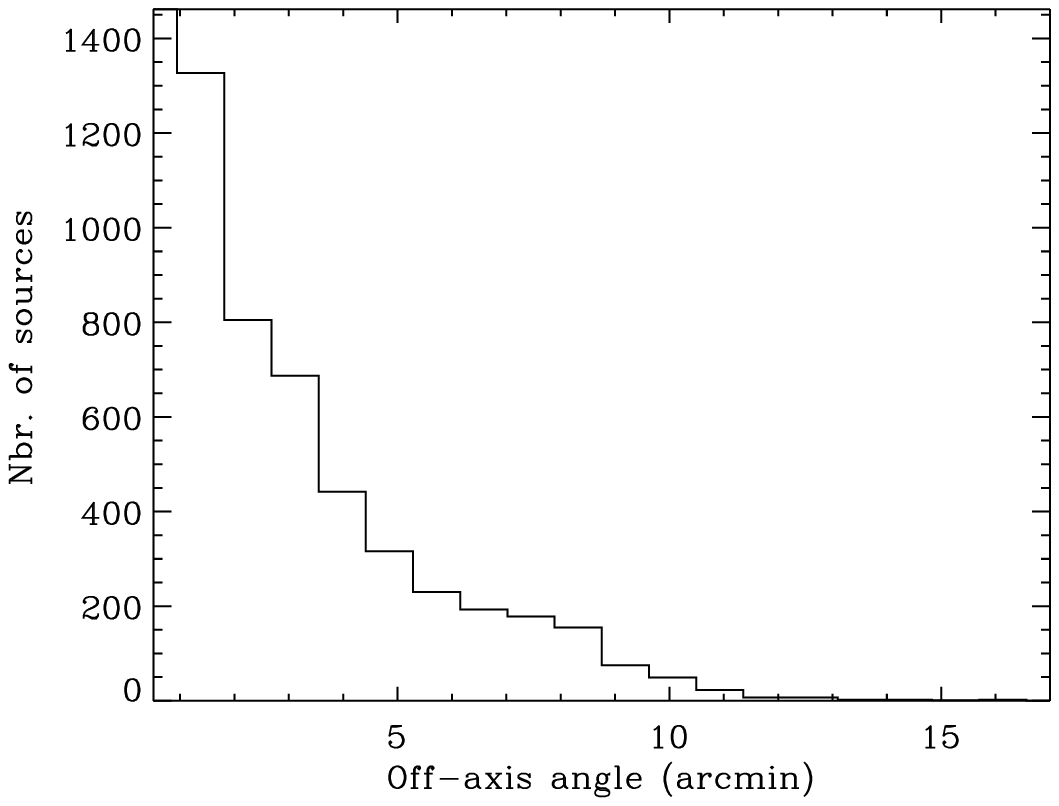}
\caption{Distribution of the off-axis angles of the EWOCS X-ray sources.}
\label{fig_theta}%
\end{figure}

Given the design of the EWOCS survey and the compact nature of Westerlund~1, it is not surprising that the majority of sources are observed at low off-axis angles, as depicted in Fig. \ref{fig_theta}. Specifically, 63.7\% of the sources (3485/5464) are located within 1 arcminute from the field center, and 87.7\% are within 3 arcminutes. Consequently, source positions are generally well-determined, with a median position error of 0.17$^{\prime\prime}$ and a 75\% quantile position error of 0.27$^{\prime\prime}$. Position errors are estimated from the single-axis standard deviations of the PSF inside the extraction region and the number of counts extracted. This precision is crucial for the search of OIR counterparts and for dynamics studies. \par

\begin{figure}[!h]
\centering
\includegraphics[width=0.5\textwidth]{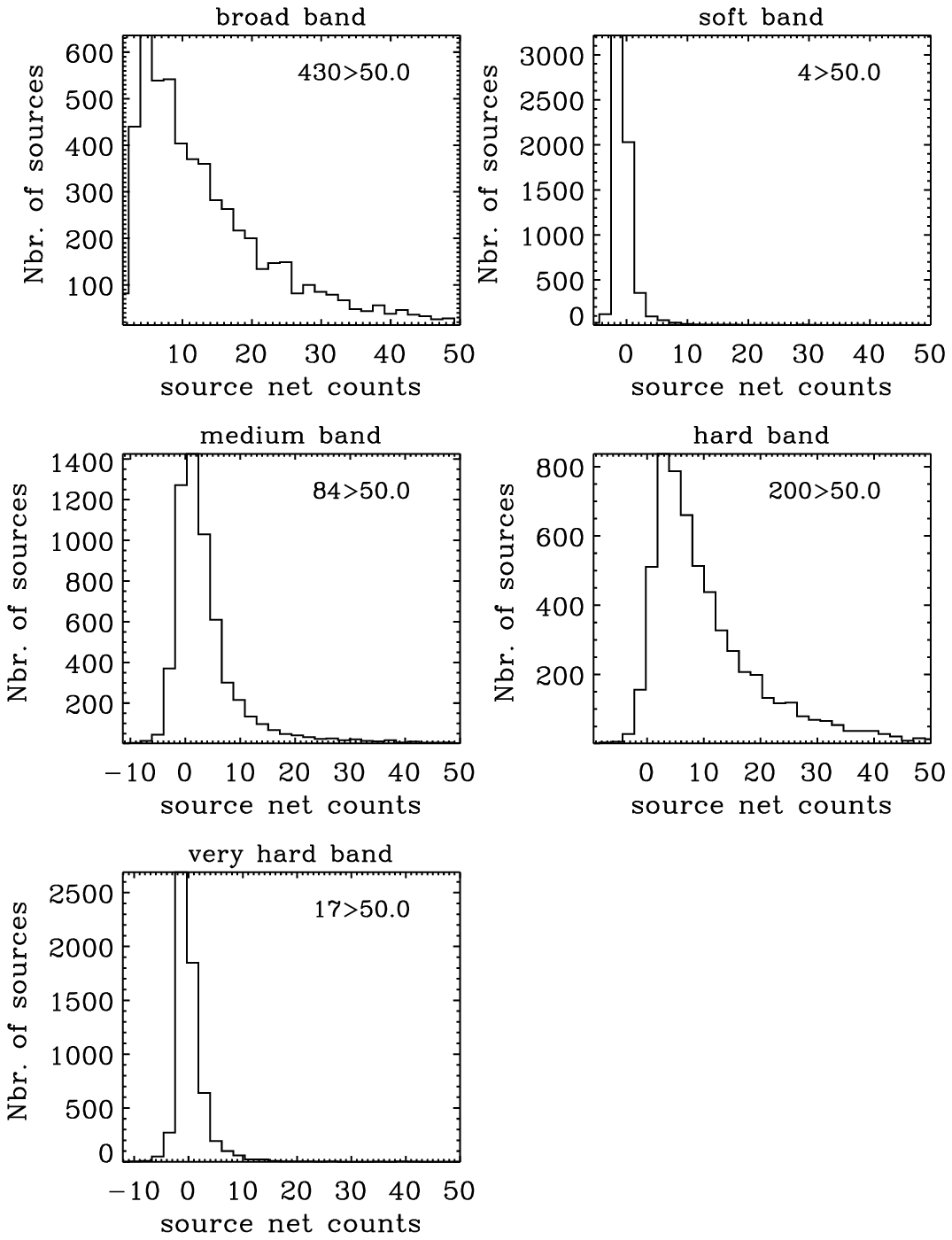}
\caption{Distributions of source net counts in the broad, soft, medium, hard, and very hard energy bands. The number of sources with more than 50 counts is indicated in the top-right corner of each panel.}
\label{fig_netcnts}%
\end{figure}

\begin{figure}[!h]
\centering
\includegraphics[width=0.48\textwidth]{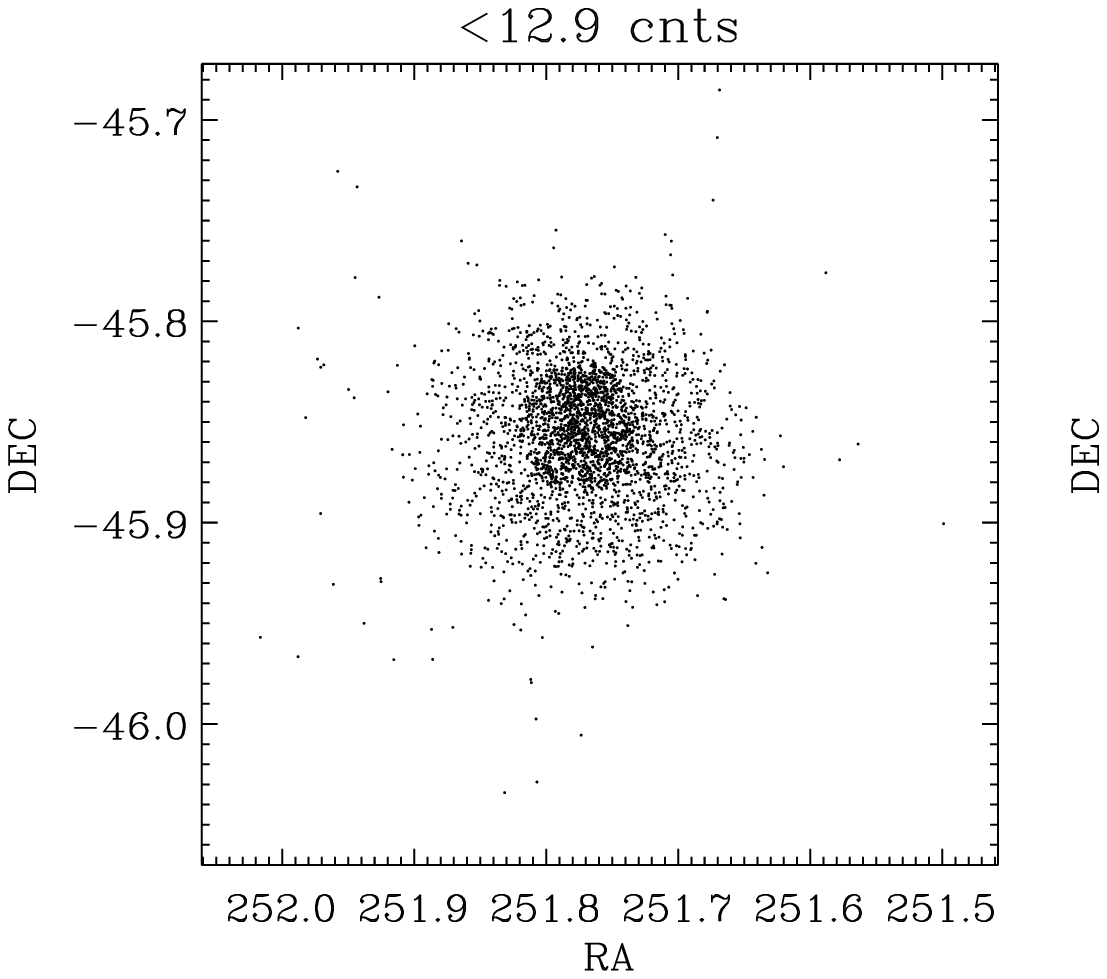}
\caption{Spatial distribution of EWOCS X-ray sources, sorted into two bins based on their net counts in the broad band.}
\label{fig_netcnts_spadis}%
\end{figure}

As depicted in Fig. \ref{fig_netcnts}, the catalog is predominantly composed of faint sources. In the broad band, the median value of the net counts is 12.9 counts. There are 607 sources (10.2\%) with fewer than 5 net counts and only 69 sources (1.2\%) with fewer than 3 counts. It is well known that the sensitivity of ACIS-I decreases with the off-axis angle. This must be taken in consideration when comparing the spatial distribution of X-ray sources detected with ACIS-I to those detected with other instruments. Fig. \ref{fig_netcnts_spadis} illustrates the spatial distributions of EWOCS X-ray sources with fewer than 12.9 net counts and those with more net counts. The former sample exhibits a higher concentration in the center of the field, with only 37 sources having an off-axis angle larger than 7$^{\prime}$. This region is considerably large compared to the size of Westerlund~1, so studies based on the spatial distribution of cluster members would not be significantly affected by the decline in sensitivity with the off-axis angle. \par

Given the design of the EWOCS observations and the intricate procedure we employed for source detection and validation, it is not currently feasible to provide a reliable estimate of catalog completeness without making strong and unverified assumptions about cluster properties, its morphology, and both mass and L$\rm_X$ distributions. Instead, we prefer to discuss the achieved completeness in future papers of this series, once the identification of OIR counterparts and the determination of true cluster members have been accomplished. In Appendix \ref{App_completeness}, however, we present a simplified analysis of completeness based on different assumptions regarding cluster morphology, along with simulations conducted using the MARX simulator.

\begin{figure}[!h]
\centering
\includegraphics[width=0.48\textwidth]{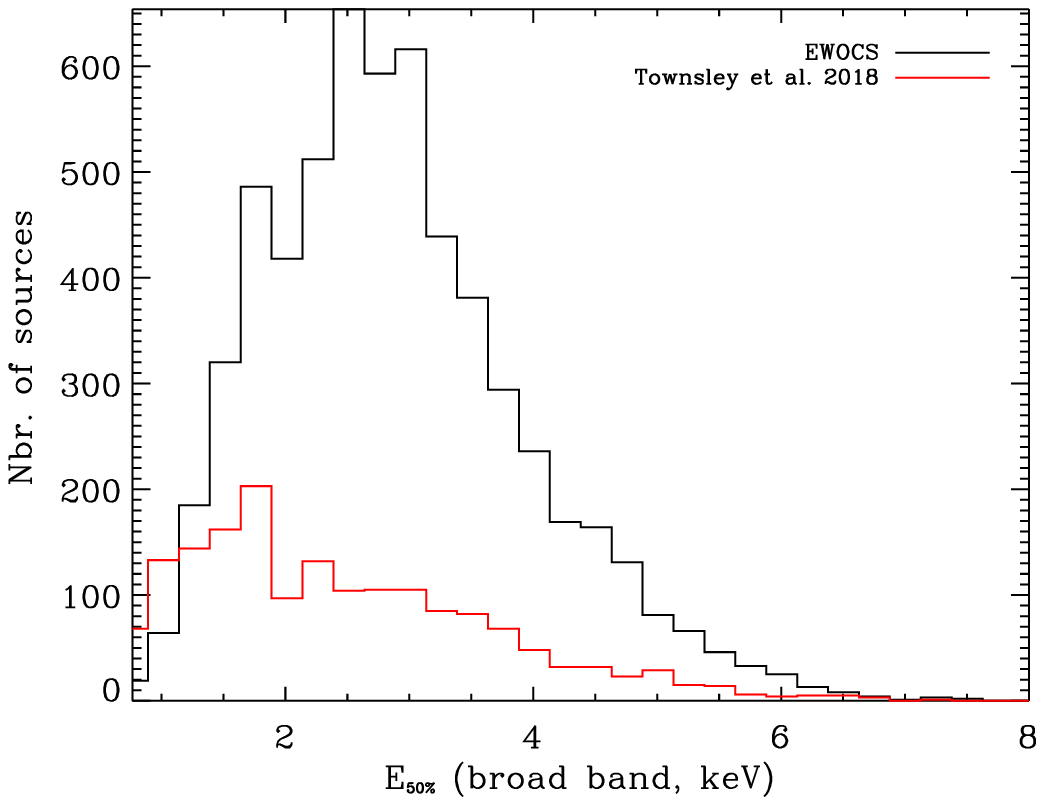}
\caption{Distribution of the median photon energy in the broad band for the validated EWOCS X-ray sources (black) and the catalog published by \citet{Townsley2018ApJS.235.43T}, in red.}
\label{fig_medE}%
\end{figure}

The distribution of the median photon energy for the EWOCS X-ray sources is shown in Fig. \ref{fig_medE}. The median value of the distribution is 2.8 keV. In the case of young stellar populations in clusters with low extinction, the median photon energy serves as a reliable indicator of membership since the coronal plasma temperature in young stars is typically higher than in older stars. However, since interstellar absorption is significant in the direction of Westerlund~1, it becomes challenging to differentiate between the absorbed background population and the young stars within the cluster. However, the secondary peak observed at energies below 2 keV in the E$\rm_{med}$ distribution could potentially be attributed to a foreground population. Nevertheless, the spatial distribution of these soft sources does not differ significantly from that of the more energetic sources. Additionally, Fig. \ref{fig_medE} includes a comparison between the photon median energy distribution of the EWOCS X-ray sources and the X-ray catalog published by \citet{Townsley2018ApJS.235.43T} based on pre-EWOCS observations, which clearly exhibits a peak below 2 keV. The evident differences in the two distributions can be due to a combination of factors: the presence of a large population of stars associated with Westerlund 1 in the EWOCS catalog (which is a factor of $\sim$5 deeper in X-ray photon flux compared to the catalog published by  \citealt{Townsley2018ApJS.235.43T}, considering the faintest sources in the two catalogs), as well as the decline in sensitivity in the soft band of the ACIS detector with the years, and the better sensitivity toward soft events of the ACIS-S detector compared with ACIS-I.\par

The catalog also includes a measure of source flux provided by AE: the photon flux (F$\rm_{photons}$), which is calculated as the ratio of the source net counts to the product of the mean effective area and nominal exposure time (thus expressed in units of photons$\,$cm$^{ -2}\,$s$^{ -1}$). A model-independent estimate of the apparent source energy flux can be calculated as 1.602$\times$10$^{-9}\times$E$\rm_{med}\times$F$\rm_{photons}$. The coefficient is derived from the conversion between keV and erg, as determined by \citet{Getman2010ApJ.708.1760G}. \par

In Appendix \ref{app-catalog}, we show ten rows of the X-ray EWOCS catalog, which is available in full at the CDS. We have also made the output table produced by AE available on the EWOCS website in its original IDL format.\footnote{https://westerlund1survey.wordpress.com/}. \par

\subsection{Specific sources}

Not surprisingly, CXOU~J16 is the brightest source in the EWOCS X-ray catalog, with a total of 71601$\pm$268 net counts collected and a photon flux of 3.45$\times$10$^{-4}\,$photons$\,$cm$^{ -2}\,$s$^{ -1}$. The source, which is strongly piled-up, is quite isolated and produces a surrounding bright background, with the closest source being at about 8 arcsec. The pulsar will be analyzed in detail in future papers of this project. \par

As explained in Sect. \ref{sect_Wd1}, Westerlund~1 hosts a unique ensemble of massive stars caught in different evolutionary stages. Understanding the mechanisms responsible for the emission of X-rays and studying both binarity and the circumstellar environment in these stars is of primary importance for EWOCS. We visually inspected the X-ray counterparts of massive stars published by \citet{Clark2020AA...635A.187C} and found 126 coincidences out of the 166 listed massive stars. The results are listed in Table \ref{tab_massive}. In the vast majority of cases, there was a clear one-to-one correspondence between the sources in the two catalogs. There are a few uncertain cases, which can be easily identified by repeated massive star IDs, EWOCS objects, or large separations. \par

The brightest X-ray massive star in the EWOCS catalog is the SgB[e] star W9 \citep{Clark2014AA...561A..15C}, with nearly 8000 net counts collected in the broad band. The intense X-ray brightness (L$\rm_X\sim$3.6$\times10^{33}\,$erg/s) and the hardness of the spectrum, as previously reported by \citet{Clark2008AA.477.147C}, are consistent with the evidence of intense mass loss rate, estimated to be around 10$^{-5}\,$M$_\odot$/yr \citep{Andrews2019AA...632A..38A}, and strong indications of binarity \citep{Ritchie2022AA...660A..89R}. W9 is the brightest source in the cluster also at radio wavelengths \citep{Andrews2019AA...632A..38A}, millimeter band \citep{Fenech2018AA...617A.137F}, and it shows very bright mid-IR emission \citep{Clark1998MNRAS.299L..43C}. \par

In terms of X-ray luminosity, W9 is followed by the post-binary blue straggler W30 \citep[O4-5Ia$^+$][]{Clark2019AA...623A..83C,Clark2008AA.477.147C}, for which a putative orbital period of approximately 6.2 days has been identified from a radial velocity series analyzed by \citet{Ritchie2022AA...660A..89R}. We have detected nearly 6000 net counts in the broad band for W30. After W9 and W30, the list of sources with net counts ranging between 190 and 5400 photons includes most of the known WR stars and OB supergiant binary systems.\par

The deep EWOCS observations provide, for the first time, candidate X-ray detections for some normal giant and subgiant stars, such as W50b and W1051. The lack of detection in the pre-EWOCS observations has been explained as a natural consequence of the lower intrinsic bolometric luminosity of these stars compared to more evolved massive stars in the cluster \citep{Clark2019AA...623A..83C}. The detection in the EWOCS observations supports this hypothesis. Faint X-ray counterparts have also been found for the two YHGs W4 and W8 (with 22 and 73 net counts, respectively), whose nature has been recently discussed by \citet{Beasor2023arXiv230316937B}, who classified them as yellow supergiants. Additionally, a faint counterpart has been detected for the BHG W1049 (with $14.16^{20.4}_{8.4}$ net counts). For the first time, faint counterparts have been found for the four O9.5II SB1 stars W1022, W1050, W1056, and W1060, as well as for the B1.5II star W1048. We also confirm the relatively faint ($75.75^{85.7}_{65.8}$ net counts) and soft (median photon energy of 1.9 keV) X-ray emission from the SB2 star (B0.5I+OB) W10, as previously reported by \citet{Clark2008AA.477.147C}. \citet{Ritchie2022AA...660A..89R} attributed the X-ray properties of this star to the possibility that the pre-EWOCS observations were made at a phase where the wind collision zone was weak or obscured. However, given the length of the EWOCS observations, it is more likely that these properties are intrinsic to the star. \par

It is quite interesting that out of the 124 sources in our catalog with more than 100 net counts, 94 do not readily match any known massive stars in the cluster. This subset will be studied in detail in future works of this series to determine their nature. It is intriguing that this sample does not seem to follow the distribution of photon energy shown in Fig. \ref{fig_medE}. In fact, its photon median energy distribution exhibits three distinct peaks: one below 2 keV (which may be dominated by foreground stars), one between 2.5 keV and 3 keV (compatible with cluster stars), and one between 3.8 keV and 4.3 keV (which could be influenced by background sources or flaring low-mass cluster members). \par   

We also provide a list of positions for the unvalidated candidate X-ray sources on the EWOCS website\footnote{https://westerlund1survey.wordpress.com/}. This list will be cross-matched with existing optical and infrared catalogs of Westerlund~1 to determine the fraction of rejected sources that could potentially be true counterparts of cluster members. Likewise, identifying optical and infrared counterparts will enable us to assess the level of contamination and the fraction of expected spurious sources in the EWOCS X-ray source catalog, as well as determine the completeness limit achieved by our survey. \par

\section{Conclusions}

In this paper, we present the EWOCS project and a new list of X-ray sources in the young supermassive star cluster Westerlund~1 and its surrounding area. The EWOCS project aims to investigate the impact of the starburst environment on the formation process of stars and planets, the dispersal of protoplanetary disks, and the evolutionary pathway of massive stars. \par

Here we present the 1$\,$Msec \emph{Chandra}/ACIS-I EWOCS observations of Westerlund~1, the workflow for data reduction, the procedure for source detection and validation, and the spectral extraction of the validated sources. Initially, we generated a preliminary list of 9420 candidate X-ray sources using the image reconstruction method, \emph{PWDetect}, \emph{WAVDETECT}, and a specific deployment of \emph{PWDetect} focused on identifying flaring stars that exhibited a significant signal above the background for a brief duration. Additionally, a few sources were manually added or obtained from existing catalogs of Westerlund~1 sources. From these input sources, we compiled the EWOCS catalog of X-ray sources in Westerlund~1 of 5963 sources successfully validated using the IDL-based software AE. \par

The median value of net counts in the EWOCS X-ray catalog is approximately 13 counts, with about 10\% of sources having fewer than 5 net counts detected in the broad energy band. The distribution of the median photon energy of the sources peaks at approximately 2.8$\,$keV, with a contribution from unrelated (foreground and background) sources that is challenging to distinguish from the candidate cluster members. The brightest source in the catalog is the magnetar CXO J164710.2-455216, with over 70000 net counts detected in the broad band. It is followed by several massive stars in Westerlund~1, including the SgB[e] star W9, the post-binary blue straggler W30, and some WR stars and supergiants in binary systems. Out of the 166 known very massive stars in Westerlund~1, we have identified a reliable X-ray counterpart for 126 of them. Additionally, we have made the first detection of an extended and rich halo surrounding the core of Westerlund~1, which will be crucial in assessing the cluster's true mass content, formation, and evolution. \par

\begin{acknowledgements}
%This paper employs a list of Chandra datasets, obtained by the Chandra X-ray Observatory, contained in \dataset[Chandra Data Collection (CDC) 153]{https://doi.org/10.25574/cdc.153} \par

We acknowledge the referee for his/her careful reading of our paper and suggestions. M.G.G. is also indebted to Leisa K. Townsley and Patrick S. Broos for their valuable suggestions and advice on data reduction and ACIS extraction. M. G. G., C. A., R. B., E. F., G. L. I., L. P., and S. S. acknowledge the INAF grant 1.05.12.05.03. K.M. acknowledges support from the Fundação para a Ciência e a Tecnologia (FCT) through the CEEC-individual contract 2022.03809.CEECIND and research grants UIDB/04434/2020 and UIDP/04434/2020. Support for this work was also provided by the National Aeronautics and Space Administration through \emph{Chandra} Proposal 21200267 issued by the \emph{Chandra} X-ray Center, which is operated by the Smithsonian Astrophysical Observatory for and on behalf of the National Aeronautics Space Administration under contract NAS8-03060. I.N. is partially supported by the Spanish Government Ministerio de Ciencia e Innovaci\'on (MCIN) and Agencia Estatal de Investigaci\'on (MCIN/AEI/10.130~39/501~100~011~033/FEDER, UE) under grant PID2021-122397NB-C22, and also by MCIN with funding from the European Union NextGenerationEU and Generalitat Valenciana in the call Programa de Planes Complementarios de I+D+i (PRTR 2022), project HIAMAS, reference ASFAE/2022/017. M. G. G. and R. B. also acknowledge partial support from the Grant INAF 2022 YODA; R. B. also from the project PRIN-INAF 2019 "Spectroscopically Tracing the Disk Dispersal Evolution". The scientific results reported in this article are based on observations made by the \emph{Chandra} X-ray Observatory. M.M.  acknowledges financial support from the European Research Council for the ERC Consolidator grant DEMOBLACK, under contract no. 770017, and from the German Excellence Strategy via the Heidelberg Cluster of Excellence (EXC 2181 - 390900948) STRUCTURES.
\end{acknowledgements}


\newpage
\addcontentsline{toc}{section}{\bf Bibliography}
\bibliographystyle{aa}
\bibliography{biblio}
%\begin{thebibliography}{}
%\end{thebibliography}

\begin{appendix}
\begin{onecolumn}

\section{Extract of the EWOCS X-ray Westerlund 1 sources catalog}
\label{app-catalog}

Ten rows of the EWOCS catalog of the X-ray sources in Westerlund 1 are shown here. The full catalog is available at the CDS.

\begin{sidewaystable*}
\setlength{\tabcolsep}{1.2mm}
\label{table_ewocscatalog}
    \centering
    \caption{Ten rows extracted from the EWOCS catalog of the X-ray sources in Westerlund~1}
   \begin{tabular}{ccccccccccccc}
   \hline
   \multicolumn{7}{c|}{Astrometry} & \multicolumn{4}{|c|}{Photometry} & \multicolumn{2}{c}{ } \\
   \hline
EWOCS-X ID & Catalog name$^{(a)}$ & $\alpha$ & $\delta$ &$\sigma_\alpha^{b}$ & $\sigma_\delta^{b}$ & $\Theta^{(c)}$ & C$\rm_t$ & C$\rm_s$ & C$\rm_m$ & C$\rm_h$ & & \\
        &                    & J2000    & J2000    & arcsec             & arcsec              & arcmin   & counts   & counts   & counts	& counts    & & \\
\hline
3001 & 164704.14-454957.4 & 251.767277 & -45.832630 & 0.06 & 0.06 & 1.6 & 38 & 0 & 11 & 27 & & \\
3002 & 164704.14-455100.2 & 251.767284 & -45.850057 & 0.06 & 0.06 & 0.8 & 32 & 1 & 11 & 20 & & \\
3003 & 164704.15-455133.6 & 251.767302 & -45.859349 & 0.05 & 0.05 & 0.6 & 47 & 0 & 13 & 34 & & \\
3004 & 164704.15-455118.1 & 251.767311 & -45.855035 & 0.05 & 0.05 & 0.6 & 50 & 0 & 11 & 39 & & \\
3005 & 164704.16-455320.2 & 251.767343 & -45.888959 & 0.09 & 0.10 & 1.9 & 14 & 1 &  5 &  8 & & \\
3006 & 164704.16-455002.9 & 251.767353 & -45.834155 & 0.10 & 0.10 & 1.6 &  7 & 0 &  2 &  5 & & \\
3007 & 164704.16-455135.0 & 251.767366 & -45.859734 & 0.06 & 0.06 & 0.6 & 19 & 0 &  5 & 14 & & \\
3008 & 164704.17-455010.5 & 251.767382 & -45.836274 & 0.08 & 0.08 & 1.4 & 23 & 0 &  5 & 18 & & \\
3009 & 164704.18-455025.7 & 251.767450 & -45.840474 & 0.06 & 0.05 & 1.2 & 36 & 1 & 13 & 22 & & \\
3010 & 164704.19-455126.4 & 251.767479 & -45.857345 & 0.07 & 0.07 & 0.6 & 27 & 0 & 10 & 17 & & \\
\hline
\hline
 & & & & & & & & & & & &  \\
 \multicolumn{13}{c}{Photometry} \\
\hline
C$\rm_{net,t}$ & C$\rm_{net,s}$ & C$\rm_{net,m}$ & C$\rm_{net,h}$ & E$_{\rm median}$ & F$\rm_{photons,t}$ & F$\rm_{photons,s}$ & F$\rm_{photons,m}$ & F$\rm_{photons,h}$ & log(P$\rm_{B,best}$) \\
counts   & counts   & counts   & counts  & keV  & photons$\,$cm$^{ -2}\,$s$^{ -1}$ & photons$\,$cm$^{ -2}\,$s$^{ -1}$ & photons$\,$cm$^{ -2}\,$s$^{ -1}$ & photons$\,$cm$^{ -2}\,$s$^{ -1}$ \\
\hline
25.9$^{+7.2}_{-6.1}$ & -0.5$^{+1.8}_{NaN} $ &  8.1$^{+4.4}_{-3.2}$ &  18.2$^{+6.3}_{-5.1}$ & 3.22 & 1.25$\times$10$^-7$ & NaN		      & 3.00$\times$10$^-8$ & 8.68$\times$10$^-8$ & -8.94 \\
12.9$^{+6.8}_{-5.7}$ &  0.7$^{+2.3}_{-0.8}$ &  3.7$^{+4.4}_{-3.3}$ &   8.4$^{+5.6}_{-4.4}$ & 2.77 & 6.97$\times$10$^-8$ & 2.32$\times$10$^-8$ & 1.50$\times$10$^-8$ & 4.49$\times$10$^-8$ & -2.23 \\
37.1$^{+7.9}_{-6.8}$ & -0.1$^{+1.8}_{NaN} $ &  8.9$^{+4.7}_{-3.5}$ &  28.2$^{+6.9}_{-5.8}$ & 2.95 & 2.19$\times$10$^-7$ & NaN		      & 3.93$\times$10$^-8$ & 1.65$\times$10$^-7$ & -15.2 \\
19.1$^{+8.3}_{-7.2}$ &  0.0$^{+1.8}_{NaN} $ & -0.1$^{+4.5}_{-3.4}$ &  19.2$^{+7.4}_{-6.3}$ & 4.15 & 8.95$\times$10$^-8$ & NaN		      & NaN		    & 8.85$\times$10$^-8$ & -5.1 \\
 7.4$^{+4.8}_{-3.7}$ &  0.7$^{+2.3}_{-0.8}$ &  3.2$^{+3.4}_{-2.1}$ &   3.4$^{+3.9}_{-2.7}$ & 2.02 & 4.15$\times$10$^-8$ & 2.14$\times$10$^-8$ & 1.36$\times$10$^-8$ & 1.90$\times$10$^-8$ & -6.61 \\
 3.1$^{+3.7}_{-2.5}$ & -0.1$^{+1.8}_{NaN} $ &  0.9$^{+2.6}_{-1.2}$ &   2.2$^{+3.4}_{-2.1}$ & 2.24 & 2.49$\times$10$^-8$ & NaN		      & 5.53$\times$10$^-9$ & 1.80$\times$10$^-8$ & -3.53 \\
 9.7$^{+5.4}_{-4.3}$ & -0.1$^{+1.8}_{NaN} $ &  1.2$^{+3.4}_{-2.1}$ &   8.6$^{+4.8}_{-3.7}$ & 3.46 & 5.56$\times$10$^-8$ & NaN		      & 5.34$\times$10$^-9$ & 4.84$\times$10$^-8$ & -2.92 \\
11.1$^{+5.9}_{-4.7}$ & -0.2$^{+1.8}_{NaN} $ &  0.9$^{+3.4}_{-2.1}$ &  10.4$^{+5.3}_{-4.2}$ & 4.99 & 5.31$\times$10$^-8$ & NaN		      & 3.62$\times$10$^-9$ & 4.87$\times$10$^-8$ & -4.68 \\
19.4$^{+7.1}_{-6.0}$ &  0.6$^{+2.3}_{-0.8}$ &  7.2$^{+4.7}_{-3.6}$ &  11.6$^{+5.8}_{-4.7}$ & 2.17 & 1.06$\times$10$^-7$ & 2.04$\times$10$^-8$ & 2.94$\times$10$^-8$ & 6.27$\times$10$^-8$ & -3.70 \\
 9.2$^{+6.3}_{-5.2}$ & -0.5$^{+1.8}_{NaN} $ &  3.7$^{+4.3}_{-3.1}$ &   6.0$^{+5.2}_{-4.1}$ & 4.16 & 4.66$\times$10$^-8$ & NaN		      & 1.44$\times$10$^-8$ & 2.98$\times$10$^-8$ & -2.60 \\
\hline
\multicolumn{13}{l}{Columns 1--11 are shown in the top table; columns 12--21 in the bottom table.} \\
\multicolumn{13}{l}{a: IAU designation.} \\
\multicolumn{13}{l}{b: single axis position error, representing only the random component of the position uncertainty.} \\
\multicolumn{13}{l}{c: Off-axis angle.} \\
\multicolumn{13}{l}{Photometric quantities are given in broad ($t$), soft ($s$), medium ($m$), and hard ($h$) bands.} \\
\multicolumn{13}{l}{C$\rm_{X}$ indicate the total counts in the X band, C$\rm_{X,net}$ the net counts. }\\
    \end{tabular}
\end{sidewaystable*}
\newpage
\section{EWOCS X-ray counterparts of the massive stars in Westerlund~1}

Table \ref{tab_massive} shows the EWOCS X-ray counterparts of the massive stars in Westerlund~1 listed in \citet{Clark2020AA...635A.187C}.

\begin{center}
\begin{longtable}{|c|c|c|c|c|c|}
\caption{ Known massive stars in the EWOCS X-ray catalog} \label{tab_massive}\\

\hline\hline
  \multicolumn{1}{|c|}{ID}    &   \multicolumn{1}{|c|}{Spectral type}         &     \multicolumn{1}{|c|}{Catalog name}   &    \multicolumn{1}{|c|}{C$\rm_{net,t}$}  & \multicolumn{1}{|c|}{F$\rm_{photons,t}$}  &\multicolumn{1}{|c|}{Sep.} \\
     \multicolumn{1}{|c|}{ }   &   \multicolumn{1}{|c|}{ }                     &   \multicolumn{1}{|c|}{ }                   & \multicolumn{1}{|c|}{counts}          & \multicolumn{1}{|c|}{photons$\,$cm$^{ -2}\,$s$^{ -1}$} &\multicolumn{1}{|c|}{arcsec}    \\
\hline
\endfirsthead
\hline\hline
  \multicolumn{1}{|c|}{ID}    &   \multicolumn{1}{|c|}{Spectral type}         &     \multicolumn{1}{|c|}{Catalog name}   &    \multicolumn{1}{|c|}{C$\rm_{net,t}$}  & \multicolumn{1}{|c|}{F$\rm_{photons,t}$}  &\multicolumn{1}{|c|}{Sep.} \\
     \multicolumn{1}{|c|}{ }   &   \multicolumn{1}{|c|}{ }                     &   \multicolumn{1}{|c|}{ }                   & \multicolumn{1}{|c|}{counts}               & \multicolumn{1}{|c|}{photons$\,$cm$^{ -2}\,$s$^{ -1}$} &\multicolumn{1}{|c|}{arcsec}    \\
\hline
\endhead
\hline \multicolumn{6}{|r|}{{continued on next page}} \\ \hline
\endfoot
\hline \hline
\endlastfoot

  W9    &  sgB[e]                &   164704.13-455031.3 &$7975.3_{7885.7}^{8065.8}$ &  3.7$\times$10$^{-5}$ &0.2 \\
  W30   &  O4-5Ia 	         &   164704.10-455039.2 &$5930.2_{5852.8}^{6008.4}$ &  2.7$\times$10$^{-5}$ &0.2 \\
  W72   &  WN7b                  &   164708.35-455045.4 &$5412.8_{5338.8}^{5487.5}$ &  4.5$\times$10$^{-5}$ &0.3 \\
  WRB   &  WN7o			 &   164705.36-455104.8 &$3046.2_{2990.6}^{3102.4}$ &  1.4$\times$10$^{-5}$ &0.1 \\
  WRU   &  WN6o                  &   164706.53-455039.1 &$1996.9_{1952.0}^{2042.3}$ &  9.6$\times$10$^{-6}$ &0.1 \\
  W44   &  WN9h                  &   164704.19-455107.2 &$1221.9_{1186.5}^{1257.6}$ &  5.6$\times$10$^{-6}$ &0.3 \\
  W239  &  WC9d                  &   164705.20-455225.0 &$903.2_{873.0}^{933.7}$    &  6.0$\times$10$^{-6}$ &0.0 \\
  W53   &  OBIa+OBIa             &   164700.38-455131.8 &$515.5_{492.4}^{538.7}$    &  2.5$\times$10$^{-6}$ &1.0 \\
  W36   &  OBIa+OBIa		 &   164705.07-455055.2 &$514.7_{491.0}^{538.7}$    &  2.4$\times$10$^{-6}$ &0.4 \\
  WRO   &  WN6o                  &   164707.65-455236.0 &$387.6_{367.7}^{407.7}$    &  2.5$\times$10$^{-6}$ &0.1 \\
  WRN   &  WC9d                  &   164659.91-455525.6 &$375.2_{355.3}^{395.2}$    &  2.2$\times$10$^{-6}$ &0.4 \\
  W27   &  O7-8Ia$^+$            &   164705.14-455041.4 &$340.1_{320.8}^{359.6}$    &  1.5$\times$10$^{-6}$ &0.1 \\
  W13   &  B0.5Ia$^+$+OB         &   164706.44-455026.1 &$258.0_{241.4}^{274.7}$    &  1.2$\times$10$^{-6}$ &0.1 \\
  WRW   &  WN6h                  &   164707.61-454922.1 &$243.0_{227.1}^{259.1}$    &  1.3$\times$10$^{-6}$ &0.3 \\
  WRJ   &  WN5h                  &   164702.47-455059.9 &$233.4_{217.5}^{249.5}$    &  1.1$\times$10$^{-6}$ &0.1 \\
  W14c  &  WN5o                  &   164706.09-455022.4 &$192.9_{178.1}^{207.8}$    &  9.2$\times$10$^{-7}$ &0.3 \\
  W24   &  O9Iab                 &   164702.15-455112.6 &$191.7_{177.2}^{206.4}$    &  9.3$\times$10$^{-7}$ &0.2 \\
  W43c  &  O9Ib                  &   164703.75-455058.5 &$188.5_{173.8}^{203.3}$    &  1.0$\times$10$^{-6}$ &0.2 \\
  1041  &  O9.5Iab               &   164704.45-455109.4 &$170.4_{156.2}^{184.8}$    &  7.9$\times$10$^{-7}$ &1.0 \\
  WRX   &  WN5o                  &   164714.13-454832.0 &$154.2_{141.3}^{167.3}$    &  8.6$\times$10$^{-7}$ &0.3 \\
  WRG   &  WN7o                  &   164704.00-455125.1 &$146.8_{133.8}^{159.9}$    &  6.8$\times$10$^{-7}$ &0.0 \\
  W50b  &  O9III                 &   164701.21-455027.6 &$136.2_{124.1}^{148.4}$    &  7.4$\times$10$^{-7}$ &1.0 \\
  W38   &  O9Iab  		 &   164702.88-455046.2 &$118.9_{106.9}^{130.9}$    &  5.6$\times$10$^{-7}$ &0.3 \\
  W37   &  O9Ib			 &   164706.01-455047.5 &$118.1_{106.1}^{130.2}$    &  5.5$\times$10$^{-7}$ &0.1 \\
  W35   &  O9Iab  		 &   164704.20-455053.7 &$110.3_{98.5}^{122.3}$	  &  5.9$\times$10$^{-7}$ &0.2 \\
  W25   &  O9Iab  		 &   164705.77-455033.4 &$108.4_{96.9}^{120.1}$	  &  5.1$\times$10$^{-7}$ &0.1 \\
  W232  &  B0Iab  		 &   164701.43-455235.2 &$104.7_{94.1}^{115.3}$	  &  6.4$\times$10$^{-7}$ &0.4 \\
  W6a   &  B0.5Iab               &   164703.04-455023.7 &$98.5_{87.9}^{109.2}$	  &  4.6$\times$10$^{-7}$ &0.1 \\
  W17   &  O9Iab                 &   164706.23-455049.3 &$96.7_{85.7}^{107.7}$	  &  4.8$\times$10$^{-7}$ &0.1 \\
  W74   &  O9.5Iab               &   164707.07-455013.0 &$93.7_{83.5}^{104.0}$	  &  4.9$\times$10$^{-7}$ &0.0 \\
  W15   &  O9Ib                  &   164706.62-455029.6 &$92.1_{81.7}^{102.5}$	  &  4.5$\times$10$^{-7}$ &0.0 \\
  W47   &  O9.5Iab               &   164702.61-455117.8 &$89.0_{78.5}^{99.5}$      &  4.4$\times$10$^{-7}$ &0.3 \\
  W57c  &  WN7o                  &   164701.59-455145.2 &$88.9_{78.9}^{98.9}$      &  5.3$\times$10$^{-7}$ &0.2 \\
  WRI   &  WN8o                  &   164700.87-455120.6 &$86.6_{76.6}^{96.6}$      &  4.0$\times$10$^{-7}$ &0.1 \\
  WRQ   &  WN6o                  &   164655.54-455134.5 &$86.6_{76.9}^{96.3}$      &  5.6$\times$10$^{-7}$ &0.4 \\
  1027  &  O9.5Iab               &   164701.02-455007.0 &$85.6_{75.8}^{95.5}$      &  4.2$\times$10$^{-7}$ &0.7 \\
  1051  &  O9III                 &   164706.98-454940.1 &$79.5_{70.0}^{88.9}$      &  4.6$\times$10$^{-7}$ &0.2 \\
  1056  &  O9.5II                &   164708.69-455101.7 &$76.3_{66.9}^{85.8}$      &  3.7$\times$10$^{-7}$ &0.6 \\
  W10   &  B0.5I+OB              &   164703.34-455034.6 &$75.8_{65.8}^{85.7}$      &  3.5$\times$10$^{-7}$ &0.3 \\
  W8a   &  F8Ia+                 &   164704.83-455025.5 &$73.7_{63.9}^{83.6}$      &  3.4$\times$10$^{-7}$ &0.8 \\
  W1    &  O9.5Iab               &   164659.39-455046.7 &$69.8_{60.9}^{78.8}$      &  3.4$\times$10$^{-7}$ &1.2 \\
  WRD   &  WN7o                  &   164706.25-455126.4 &$69.3_{59.8}^{78.8}$      &  3.1$\times$10$^{-7}$ &0.1 \\
  W62a  &  B0.5Ib                &   164702.52-455138.0 &$69.1_{60.0}^{78.1}$      &  4.0$\times$10$^{-7}$ &0.2 \\
  W65   &  O9Ib                  &   164703.88-455146.5 &$67.9_{58.9}^{76.9}$      &  3.7$\times$10$^{-7}$ &0.2 \\
  WRV   &  WN8o                  &   164703.79-455038.7 &$66.8_{57.8}^{75.9}$      &  4.9$\times$10$^{-7}$ &0.1 \\
  1037  &  O9.5II                &   164702.84-455006.4 &$64.8_{56.1}^{73.3}$      &  3.2$\times$10$^{-7}$ &0.1 \\
  W28   &  B2Ia                  &   164704.66-455038.5 &$63.2_{53.1}^{73.3}$      &  2.9$\times$10$^{-7}$ &0.1 \\
  W61b  &  O9.5Iab               &   164702.56-455141.9 &$61.3_{52.5}^{70.0}$      &  3.1$\times$10$^{-7}$ &0.3 \\
  1030  &  O9.5Iab               &   164701.67-455258.0 &$60.1_{51.9}^{68.4}$      &  3.4$\times$10$^{-7}$ &0.3 \\
  1040  &  O9-9.5I-III           &   164704.59-455008.1 &$59.9_{51.7}^{68.0}$      &  4.0$\times$10$^{-7}$ &1.0 \\
  1061  &  O9-9.5III             &   164709.61-455040.4 &$59.7_{51.3}^{68.1}$      &  3.1$\times$10$^{-7}$ &1.3 \\
  W84   &  O9.5Ib                &   164659.03-455028.3 &$57.0_{49.1}^{64.8}$      &  4.4$\times$10$^{-7}$ &0.1 \\
  1064  &  O9.5Iab               &   164711.50-455000.0 &$56.4_{48.3}^{64.5}$      &  2.9$\times$10$^{-7}$ &0.6 \\
  W241  &  WC9                   &   164705.96-455208.3 &$56.3_{48.2}^{64.3}$      &  3.9$\times$10$^{-7}$ &0.9 \\
  1060  &  O9.5II                &   164709.19-455048.4 &$56.2_{47.8}^{64.5}$      &  2.8$\times$10$^{-7}$ &0.1 \\
  1036  &  O9.5Ia                &   164702.78-455212.7 &$55.8_{47.8}^{63.8}$      &  3.8$\times$10$^{-7}$ &0.3 \\
  1004  &  OeBe star             &   164653.44-455300.3 &$53.9_{45.9}^{61.8}$      &  2.7$\times$10$^{-7}$ &0.8 \\
  1058  &  O9III                 &   164708.89-455124.5 &$53.7_{45.8}^{61.6}$      &  3.3$\times$10$^{-7}$ &0.1 \\
  W56b  &  O9.5Ib                &   164658.87-455145.9 &$52.5_{44.9}^{60.1}$      &  4.8$\times$10$^{-7}$ &0.2 \\
  W29   &  O9Ib                  &   164704.40-455039.9 &$51.5_{43.5}^{59.5}$      &  3.7$\times$10$^{-7}$ &0.1 \\
  1023  &  O9III                 &   164700.14-455110.3 &$49.7_{42.0}^{57.8}$      &  2.3$\times$10$^{-7}$ &1.0 \\
  W53   &  OBIa+OBIa             &   164700.55-455132.0 &$47.6_{39.4}^{56.3}$      &  2.3$\times$10$^{-7}$ &0.7 \\
  1034  &  O9.5Iab               &   164702.52-455148.3 &$47.4_{40.1}^{55.3}$      &  2.9$\times$10$^{-7}$ &0.1 \\
  1063  &  O9III                 &   164710.74-454947.8 &$46.8_{39.5}^{54.6}$      &  2.3$\times$10$^{-7}$ &0.6 \\
  1005  &  B0Iab                 &   164654.28-455154.8 &$43.8_{37.0}^{51.3}$      &  2.7$\times$10$^{-7}$ &0.4 \\
  1047  &  O9.5II                &   164706.12-455232.2 &$43.4_{36.4}^{51.0}$      &  2.9$\times$10$^{-7}$ &0.2 \\
  W41   &  O9Iab                 &   164702.70-455057.1 &$43.0_{35.7}^{50.8}$      &  2.3$\times$10$^{-7}$ &0.2 \\
  1033  &  O9-9.5I-III           &   164702.37-455234.2 &$42.8_{36.1}^{50.1}$      &  2.7$\times$10$^{-7}$ &0.2 \\
  1018  &  O9.5Iab               &   164658.28-455057.0 &$41.8_{35.1}^{49.1}$      &  2.5$\times$10$^{-7}$ &0.4 \\
  W11   &  B2                    &   164702.24-455046.8 &$41.4_{34.1}^{49.2}$      &  1.9$\times$10$^{-7}$ &0.2 \\
  1040  &  O9-9.5I-III           &   164704.54-455009.0 &$36.0_{29.8}^{42.8}$      &  2.4$\times$10$^{-7}$ &0.3 \\
  1038  &  O9III                 &   164703.49-454857.1 &$34.8_{28.3}^{41.9}$      &  1.7$\times$10$^{-7}$ &1.1 \\
  1007  &  O9-9.5III             &   164654.90-455005.8 &$34.5_{28.3}^{41.2}$      &  2.1$\times$10$^{-7}$ &0.5 \\
  W243  &  LBV                   &   164707.50-455229.0 &$33.7_{27.5}^{40.4}$      &  2.4$\times$10$^{-7}$ &0.7 \\
  1043  &  O9.5II-III		 &   164704.56-455059.5 &$32.3_{25.8}^{39.2}$      &  2.2$\times$10$^{-7}$ &0.2 \\
  W86   &  O9.5Ib 		 &   164657.15-455010.0 &$30.3_{24.6}^{36.5}$      &  1.8$\times$10$^{-7}$ &0.1 \\
  W61a  &  B0.5Ia  		 &   164702.27-455141.7 &$28.6_{22.0}^{35.6}$      &  1.5$\times$10$^{-7}$ &0.2 \\
  W46b  &  O9.5Ib 		 &   164703.67-455120.5 &$28.5_{21.4}^{36.2}$      &  1.3$\times$10$^{-7}$ &0.9 \\
  1066  &  O9III                 &   164712.60-455055.6 &$28.3_{22.8}^{34.4}$      &  2.0$\times$10$^{-7}$ &1.2 \\
  1050  &  O9.5II                &   164706.77-454955.2 &$26.4_{20.8}^{32.6}$      &  1.3$\times$10$^{-7}$ &0.0 \\
  WRH   &  WC9d                  &   164704.23-455120.2 &$26.3_{19.3}^{33.9}$      &  1.2$\times$10$^{-7}$ &0.1 \\
  1029  &  O9-9.5III             &   164701.50-454950.1 &$25.5_{19.9}^{31.6}$      &  1.2$\times$10$^{-7}$ &0.6 \\
  W46a  &  B1Ia                  &   164703.90-455119.9 &$24.5_{17.2}^{32.4}$      &  1.1$\times$10$^{-7}$ &0.4 \\
  W21   &  B0.5Ia                &   164701.10-455113.7 &$24.4_{18.3}^{31.0}$      &  1.1$\times$10$^{-7}$ &0.1 \\
  W5    &  WN10/B0.5Ia+WRS       &   164702.98-455018.5 &$23.7_{18.2}^{29.7}$      &  1.2$\times$10$^{-7}$ &1.0 \\
  1055  &  B0Ib(+O?)		 &   164707.82-455147.1 &$23.1_{17.8}^{28.8}$      &  1.9$\times$10$^{-7}$ &1.2 \\
  W4    &  F3Ia$^+$  		 &   164701.54-455037.1 &$22.3_{16.6}^{28.6}$      &  1.0$\times$10$^{-7}$ &1.3 \\
  1065  &  B0Ib			 &   164711.60-454922.6 &$22.2_{16.8}^{28.2}$      &  1.1$\times$10$^{-7}$ &0.2 \\
  1048  &  B1.5                  &   164706.28-455104.0 &$21.4_{16.5}^{26.9}$      &  1.7$\times$10$^{-7}$ &0.3 \\
  W34   &  B0Ia			 &   164704.39-455047.3 &$21.3_{13.6}^{29.4}$      &  1.0$\times$10$^{-7}$ &0.1 \\
  W228b &  O9Ib			 &   164658.13-455301.2 &$21.1_{16.4}^{26.5}$      &  1.5$\times$10$^{-7}$ &0.9 \\
  1059  &  O9III? 		 &   164709.08-455320.7 &$21.1_{16.3}^{26.4}$      &  1.4$\times$10$^{-7}$ &0.3 \\
  W43c  &  O9Ib			 &   164703.70-455057.7 &$21.0_{15.6}^{27.0}$      &  2.1$\times$10$^{-7}$ &0.8 \\
  1044  &  O9-9.5III             &   164705.56-454951.8 &$19.8_{14.6}^{25.5}$      &  1.0$\times$10$^{-7}$ &0.3 \\
  W43b  &  B1Ia                  &   164703.52-455056.6 &$19.8_{13.2}^{26.8}$      &  1.1$\times$10$^{-7}$ &0.1 \\
  1059  &  O9III?                &   164709.11-455319.4 &$18.7_{14.2}^{23.7}$      &  1.4$\times$10$^{-7}$ &1.3 \\
  1042  &  O9.5II                &   164704.66-455206.8 &$17.9_{13.0}^{23.3}$      &  1.3$\times$10$^{-7}$ &1.1 \\
  W2a   &  B2Ia                  &   164659.77-455051.8 &$17.3_{12.3}^{22.8}$      &  8.1$\times$10$^{-8}$ &0.9 \\
  1024  &  O9.5Iab               &   164700.78-455102.0 &$16.6_{11.8}^{21.9}$      &  8.6$\times$10$^{-8}$ &0.6 \\
  W50b  &  O9III                 &   164701.11-455026.6 &$16.5_{11.5}^{22.0}$      &  8.7$\times$10$^{-8}$ &0.7 \\
  W228b &  O9Ib                  &   164658.02-455301.1 &$16.3_{12.0}^{21.0}$      &  1.2$\times$10$^{-7}$ &0.3 \\
  W243  &  LBV                   &   164707.62-455228.4 &$15.9_{11.4}^{21.1}$      &  1.2$\times$10$^{-7}$ &0.7 \\
  W1    &  O9.5Iab               &   164659.20-455045.4 &$15.8_{11.0}^{21.2}$      &  7.9$\times$10$^{-8}$ &1.4 \\
  W4    &  F3Ia+                 &   164701.35-455036.5 &$15.6_{10.5}^{21.2}$      &  7.4$\times$10$^{-8}$ &0.8 \\
  1032  &  O9-9.5III	         &   164702.32-455017.1 &$15.3_{10.4}^{20.7}$      &  7.3$\times$10$^{-8}$ &0.7 \\
  1016  &  O9-9.5III	         &   164658.09-455247.1 &$15.2_{10.7}^{20.2}$      &  8.2$\times$10$^{-8}$ &0.2 \\
  W54   &  B0.5Iab               &   164703.14-455131.2 &$14.7_{9.30}^{20.7}$      &  6.9$\times$10$^{-8}$ &1.2 \\
  1014  &  O9-9.5III             &   164657.81-455119.3 &$14.4_{10.1}^{19.3}$      &  8.7$\times$10$^{-8}$ &0.4 \\
  1010  &  O+O?                  &   164655.99-455210.1 &$14.4_{10.2}^{19.2}$      &  8.7$\times$10$^{-8}$ &0.7 \\
  1015  &  O9III                 &   164657.97-455141.0 &$14.2_{10.3}^{18.6}$      &  1.5$\times$10$^{-7}$ &0.3 \\
  1049  &  B1-2Ia+               &   164706.66-454738.8 &$14.2_{8.4}^{20.4}$   	  &  8.5$\times$10$^{-8}$ &0.3 \\
  1031  &  O9III                 &   164701.90-455056.1 &$14.0_{9.8}^{18.6}$   	  &  9.9$\times$10$^{-8}$ &0.2 \\
  1043  &  O9.5II-III            &   164704.63-455059.4 &$13.1_{8.6}^{18.1}$   	  &  1.2$\times$10$^{-7}$ &0.7 \\
  W23a  &  B2Ia+BI?		 &   164702.56-455108.8 &$12.9_{7.1}^{19.2}$   	  &  6.0$\times$10$^{-8}$ &0.1 \\
  W63a  &  B0Iab  		 &   164703.41-455157.4 &$12.7_{8.3}^{17.6}$   	  &  9.0$\times$10$^{-8}$ &0.3 \\
  W55   &  B0Ia			 &   164658.40-455131.1 &$12.5_{8.5}^{17.0}$   	  &  8.9$\times$10$^{-8}$ &0.0 \\
  1012  &  O9-9.5III             &   164656.95-455055.6 &$12.4_{8.3}^{17.1}$   	  &  7.1$\times$10$^{-8}$ &0.3 \\
  W238  &  B1Iab                 &   164704.41-455227.7 &$12.1_{7.8}^{17.1}$   	  &  7.5$\times$10$^{-8}$ &0.1 \\
  1046  &  O+O?                  &   164705.98-454955.4 &$11.6_{7.3}^{16.6}$   	  &  5.6$\times$10$^{-8}$ &1.4 \\
  1045  &  O9.5II                &   164705.83-455155.1 &$11.6_{8.0}^{15.8}$   	  &  1.0$\times$10$^{-7}$ &0.2 \\
  W75   &  M4Ia                  &   164708.96-454958.7 &$11.6_{7.3}^{16.4}$   	  &  6.3$\times$10$^{-8}$ &0.4 \\
  1021  &  O9-9.5III             &   164658.77-455432.0 &$11.5_{6.8}^{16.8}$   	  &  6.2$\times$10$^{-8}$ &0.1 \\
  1013  &  O+O?                  &   164657.54-455231.0 &$11.5_{7.4}^{16.2}$   	  &  6.4$\times$10$^{-8}$ &0.6 \\
  1035  &  O9-9.5III             &   164702.67-455151.2 &$11.3_{7.1}^{16.1}$   	  &  7.6$\times$10$^{-8}$ &0.4 \\
  1046  &  O+O?                  &   164706.09-454957.7 &$11.1_{6.8}^{15.9}$   	  &  5.4$\times$10$^{-8}$ &1.3 \\
  1017  &  O9-9.5III             &   164658.24-455033.8 &$10.8_{6.7}^{15.5}$   	  &  6.0$\times$10$^{-8}$ &0.1 \\
  W20   &  M5Ia                  &   164703.11-455218.9 &$8.8_{5.1}^{13.3}$   	  &  5.4$\times$10$^{-8}$ &0.3 \\
  1028  &  O9-9.5                &   164701.32-455137.5 &$8.7_{4.5}^{13.6}$   	  &  4.5$\times$10$^{-8}$ &0.6 \\
  1054  &  O9-9.5                &   164707.64-455141.1 &$8.0_{5.0}^{11.7}$   	  &  8.3$\times$10$^{-8}$ &0.3 \\
  W78   &  B1Ia			 &   164701.48-454957.4 &$7.8_{4.0}^{12.2}$   	  &  3.8$\times$10$^{-8}$ &0.6 \\
  W373  &  B0Iab  		 &   164657.72-455320.0 &$7.7_{4.1}^{11.9}$   	  &  4.6$\times$10$^{-8}$ &0.1 \\
  1026  &  O9-9.5III             &   164701.01-454948.8 &$7.0_{3.3}^{11.3}$   	  &  3.5$\times$10$^{-8}$ &0.5 \\
  W71   &  B2.5Ia                &   164708.57-455049.8 &$5.5_{2.1}^{9.4}$   	  &  4.5$\times$10$^{-8}$ &1.5 \\
  1020  &  O9-9.5+O?             &   164658.49-455228.4 &$5.2_{2.5}^{8.5}$   	  &  3.7$\times$10$^{-8}$ &1.3 \\
  1008  &  O9.5II                &   164655.45-455154.2 &$5.1_{2.1}^{8.7}$   	  &  3.3$\times$10$^{-8}$ &0.1 \\
  1062  &  O+O?                  &   164710.65-455047.2 &$5.0_{1.9}^{8.7}$   	  &  3.2$\times$10$^{-8}$ &0.9 \\
  1022  &  O9.5II                &   164659.88-455025.1 &$4.6_{1.6}^{8.1}$   	  &  3.0$\times$10$^{-8}$ &0.5 \\
  1045  &  O9.5II                &   164705.86-455154.2 &$4.4_{2.1}^{7.5}$   	  &  4.8$\times$10$^{-8}$ &0.8 \\
  W29   &  O9Ib                  &   164704.47-455039.5 &$4.3_{1.4}^{7.8}$   	  &  4.8$\times$10$^{-8}$ &0.7 \\

\end{longtable}
\end{center}
\end{onecolumn}

\newpage 
\begin{twocolumn}

\section{Estimate of catalog completeness}
\label{App_completeness}

The resulting completeness of our survey depends not only on the total exposure, but also on source crowding and the bright and irregular background. A full understanding of completeness will only be possible after the identification and classification of the OIR counterparts of the X-ray sources in order to distinguish between cluster members and sources in the foreground and background. However, we conducted some simple simulations using MARX, which, despite being based on strong assumptions, can provide some hints about completeness.\par

In order to simulate the cluster population, since the true shape of the IMF of Westerlund~1 is still a subject of debate, particularly in the low-mass regime, we made the assumption that the cluster IMF follows the law proposed by \citet{Kroupa01}, which is applicable to most known young stellar clusters. We understand that the starburst environment can influence the distribution of stellar masses, leading to different mass functions. However, at this level of approximation, this is considered a secondary effect. To accommodate the compact morphology of the cluster, we assumed that cluster members are distributed according to a Gaussian function with a full width at half maximum of 4 arcminutes. Therefore, we did not account for the asymmetric morphology of Westerlund~1, as suggested by previous authors \citep[e.g.,][]{Gennaro2011}. Additionally, we assumed a total cluster mass of 45000 solar masses, encompassing stars with masses as low as 0.08 solar masses. \par

To convert the mass distribution into an L$\rm_X$ distribution, we utilized the L$\rm_X$ versus mass distribution derived from the \emph{Chandra} Orion Ultradeep Project (COUP) conducted in the Orion Nebula Cluster \citep{PreibischFeigelson2005}, accounting for its observed spread. We chose this distribution because the COUP survey provides the most complete X-ray observation of a young stellar cluster. However, it should be noted that this distribution may not accurately represent the population of Westerlund~1 due to differences in age and the presence of a distinct massive stellar population in this cluster. To account for this massive stellar population, we simply added the massive sources identified by \citet{Clark2005A&A...434..949C} with their corresponding measured L$\rm_X$ values to the simulated cluster population. Additionally, we normalized the COUP L$\rm_X$ versus mass distribution to account for the decline in stellar X-ray luminosity with age \citep{PreibischFeigelson2005}, and we used the specified values for cluster distance and absorption to convert luminosity into flux. \par

We simulated a 1 Msec ACIS-I observation of this fake cluster, taking into account instrumental background\footnote{https://cxc.harvard.edu/cal/Acis/detailed\_info.html}, and performed source detection using \emph{Wavdetect} (thus not accounting for the source validation procedure we adopted with AE). By comparing the input and output lists of sources, we determined that the completeness in the 0.8-2 solar mass range is approximately 40\% within the central 4 arcminute region, decreasing by approximately 10\% in the inner 1 arcminute region. For more massive stars, the estimated completeness is around 85\% regardless of the distance from the cluster center. It is important to note that this is a preliminary estimation of the completeness of the EWOCS X-ray catalog, which will be further validated through the identification of OIR counterparts and source classification.\par
\end{twocolumn}

\end{appendix}
\end{document}